\documentclass[twocolumn,aps,prc,showpacs,showkeys,amsfonts,amssymb,amsmath,mathrsfs]{revtex4}
\usepackage{graphicx}

\usepackage[english]{babel}
\newcommand{\ket}[1]{\mathinner{|{#1}\rangle}}
\newcommand{\bra}[1]{\mathinner{\langle{#1}|}}

\newcommand{\EXP}[1]{\mathrm{e}^{#1}}
 \newcommand{\BIGEXP}[1]{\text{\Large e}^{#1}}
\DeclareMathOperator{\ch}{ch}
\DeclareMathOperator{\sh}{sh}
\DeclareMathOperator{\re}{Re}
\DeclareMathOperator{\im}{Im}
\DeclareMathOperator*{\tr}{tr}

\newcommand{\imat}{{\mathrm{i}}}
\newcommand{\dmat}{\mathrm{d}}

\newcommand{\DEF}{\overset{\mathrm{def}}{=}}
\newcommand{\DEFt}{\;\smash{\overset{\text{\tiny def}}{=}}}
\newcommand{\scl}{\underset{\hbar\to0}{\sim}}

\newcommand{\opS}{\hat{\mathsf{S}}}

\newcommand{\goe}{\mathfrak{e}}
\newcommand{\gol}{\mathfrak{l}}
\newcommand{\gor}{\mathfrak{r}}
\newcommand{\goc}{\mathfrak{c}}
\newcommand{\gom}{\mathfrak{m}}
\newcommand{\goo}{\mathfrak{o}}

\newcommand{\gos}{\mathfrak{s}}
\newcommand{\goS}{\mathfrak{S}}

\begin{document}

\author{J\'er\'emy Le Deunff and Amaury Mouchet}
\email[email: ]{jeremy.ledeunff@lmpt.univ-tours.fr, mouchet@lmpt.phys.univ-tours.fr}
\affiliation{Laboratoire de Math\'ematiques
  et de Physique Th\'eorique, Universit\'e Fran\c{c}ois Rabelais de Tours --- \textsc{\textsc{cnrs (umr 6083)}},
F\'ed\'eration Denis Poisson,
 Parc de Grandmont 37200
  Tours,  France.}

\title{Instantons revisited: dynamical tunnelling and resonant tunnelling }
\date{\today}

\begin{abstract}
Starting from trace formulae for the tunnelling splittings (or decay rates) analytically continued in the complex
time domain, we obtain
explicit semiclassical expansions in terms of complex trajectories that are selected 
with appropriate complex-time paths. We show how this instanton-like approach, which takes advantage of
an incomplete Wick rotation, accurately reproduces
tunnelling effects not only in the usual double-well potential but also in situations where 
a pure Wick rotation is insufficient, for instance dynamical tunnelling or resonant tunnelling.
Even though only one-dimensional autonomous Hamiltonian systems are quantitatively studied,
we discuss the relevance of our method for multidimensional and/or chaotic tunnelling. 
\end{abstract}


\maketitle

\section{Introduction: events occur in complex time}\label{sec:introduction}

Instantons generally refer to solutions of classical equations in the
Euclidean space-time, \textit{i.e.}  once a Wick rotation~$t\mapsto-\imat t$ has
been performed on time~$t$ in the Minkowskian space-time.  Since the
mid-seventies, they have been extensively used in gauge field theories
 to describe tunnelling between degenerate
vacua~\cite{Shifman94a}. In introductory texts~\cite[for instance]{Coleman85a,ZinnJustin02a}, 
they are first presented within
the framework of quantum mechanics: when the classical Hamiltonian of a system
with one degree of freedom has the usual form
\begin{equation}\label{eq:Hp2plusV}
H(p,q)=\frac{1}{2}\,p^2+V(q) \end{equation}
 ($p$ and
$q$ denote the canonically conjugate variables), the Wick rotation
induces an inversion of the potential and, then, some classical real
solutions driven by the transformed Hamiltonian $p^2/2-V(q)$ can be
exploited to quantitatively describe a tunnelling transition. As far as we know, in
this context, only the simplest situations have been considered,
namely the tunnelling decay from an isolated minimum of~$V$ to a
continuum and the tunnelling oscillations between  $N$ degenerated
minima of $V$ that are related by an $N$-fold symmetry. In those cases, what can be captured is
tunnelling at the lowest energy only.
However, not to speak of the highly non-trivial cases of tunnelling in non-autonomous and/or
non-separable multidimensional 
systems, there are many situations that cannot be straightforwardly
  treated with a simple inversion of the 1d, time-independent,
  potential.

First, tunnelling --- \textit{i.e.} any quantum phenomenon that cannot be
described by \emph{real} classical solutions of the original 
(non Wick-rotated) Hamilton's equations --- may manifest itself through
a transition that is not necessarily a classically forbidden jump in
\emph{position}~\cite{Davis/Heller81b}. For instance, the reflection
above an energy barrier, as a forbidden jump in momentum, is indeed a
tunnelling process~\cite{Maitra/Heller96a,Maitra/Heller97b}.  In the following, we will
consider the case of a simple pendulum whose dynamics is governed by
the potential~$V(q)=\gamma\cos q$; for energies larger than the
strength~$\gamma>0$ of the potential, one can observe a quantum
transition between states rotating in opposite direction while two
distinct rotational classical solutions, obtained one from each other
by the reflection symmetry, are always disconnected in real
phase-space.

A second example is provided by a typical situation of resonant
tunnelling: when, for instance, $V$ has a third, deeper, well which lies in
between two symmetric wells (see Fig.~\ref{fig:3puits}a), the oscillation frequency
between the latter can be affected by several orders of magnitude, when
two eigenenergies get nearly degenerate with a third one, corresponding
to a state localised in the central well. Then, we lose the customary
exponential weakness of tunnelling and it is worth stressing, coming back for a second to
quantum field theory, that having a nearly full tunnelling transmission
through a double barrier may have drastic consequences in some
cosmological models.  In this example, one can immediately see (Fig.~\ref{fig:3puits}b) that
$-V$ also  has an energy barrier and working with a complete Wick rotation
 only remains insufficient 
in that case. 

In order to describe the tunnelling transmission 
at an energy~$E$ below the top of an energy barrier,
which may be crucial in some
chemical reactions, the pioneer works by
Freed~\cite{Freed72a}, George and Miller~\cite{George/Miller72a, Miller74a}, 
have shown that
the computation of a Green's function~$\tilde{G}(q_f,q_i;E)$ (or a scattering matrix element)
 requires taking into consideration classical trajectories with a complex time.
 These complex times come out when looking for 
the saddle-point main contributions to  the Fourier transform of the time propagator,
\begin{equation}\label{eq:GE}
  \tilde{G}(q_f,q_i;E)=\frac{1}{\sqrt{2\pi\hbar}}\int_0^\infty G(q_f,q_i;t)\,\EXP{\imat E t/\hbar}\;\dmat t\;,
\end{equation}
 which is, up to now,
  the common step shared by all the approaches involving complex time \cite[for instance]{Weiss/Haeffner83a,Carlitz/Nicole85a,Ilgenfritz/Perlt92a,Maitra/Heller97a,Creagh/Whelan99a}.
Though well-suited for the study of scattering, indirect computations are required
to extract from the poles of the energy Green's function~\eqref{eq:GE}
 some spectral signatures of tunnelling in bounded systems. Generically, these 
signatures appear
as small splittings between two quasi-degenerate energy levels and can be seen as a narrow avoided
crossing of the two levels when a classical parameter is varied~\cite{Ozorio84a,Farrelly/Uzer86a,Heller95a,Tomsovic98b}. 
In the present article, we propose a unified treatment that provides a direct computation of these splittings
 (formulae~\ref{eq:DeltaE0} or~\ref{eq:DeltaEn});
as shown in section~\ref{sec:starting points}, it  takes full advantage
of the possibility of working not necessarily with purely imaginary
time, but with a general parametrisation of complex time as first suggested in~\cite{Mclaughlin72a}.
 The 
semiclassical approach naturally follows (sec.~\ref{sec:semiclassical}) and some
general asymptotic expansions can be written (equations~\eqref{eq:scDeltangeneral} and~\eqref{eq:scDeltan})
and simplified (equation \eqref{eq:scDeltan_Nwells}); 
they constitute the main results of this paper.
To understand where these formulae come from and  how they work, we will start
with the paradigmatic case of the
double-well potential (sec.~\ref{sec:doublewell}) and  the simple pendulum
(sec.\ref{sec:pendulum}); we defer some general and technical justifications
in the appendices. Then
we will treat the resonant case in
detail in sec.~\ref{sec:resonanttunnelling}, where an appropriate incomplete Wick
rotation~$t\mapsto\EXP{-\imat\theta} t$ provides the key to showing
how interference
effects \`a la Fabry-P\'erot between several complex trajectories
reproduce the non-exponential behaviour of resonant
tunnelling, already at work in open systems with a double barrier \cite{Bohm51a, Zohta90a}.  
After having shown how to adapt our method 
to the computations of escape rates from a stable island in phase-space (sec.~\ref{sec:escape_rate}),
we will conclude with more long--term considerations by
explaining how our approach provides a
natural and new starting point for studying tunnelling in
multidimensional systems.

\section{Tunnelling splittings}\label{sec:starting points}

A particularly simple signature of tunnelling
can be identified when the Hamiltonian has a two fold symmetry and, therefore, 
we will consider quantum systems whose time-independent
Hamiltonian~$\hat{H}=H(\hat{p},\hat{q})$ commute with an
operator~$\opS$ such that~$\opS^2=1$ (the~$\ \hat{}\ $ allows to distinguish 
the quantum operators from the classical phase-space functions or maps). 
In most cases, $\opS$  stands for the parity operator:
\begin{equation}\label{eq:Hparity}
   H(-p,-q)=H(p,q)\;.
\end{equation}
 The
spectrum of $\hat{H}$ can be classified according to $\opS$ and,
for simplicity, we will always consider a bounded system whose discrete
energy spectrum and the associated orthonormal eigenbasis are defined by
\begin{equation}
   \hat{H}\ket{\phi_n^\pm} = E_{n}^{\pm}\ket{\phi_n^\pm}\;;\quad 
   \opS\ket{\phi_n^\pm} = \pm\ket{\phi_n^\pm}
\end{equation}
where~$n$ is a natural integer. When the Planck constant~$\hbar$ is small compared
to the typical classical actions, standard semi-classical 
analysis~\cite{Takahashi/Saito85a,Heller89a,TorresVega/Frederick90a} shows
that one can associate some classical regions in phase-space to each eigenstate~$\ket{\phi_n^\pm}$.
This can be done by constructing a phase-space representation of~$\ket{\phi_n^\pm}$, typically
the Wigner or the Husimi representation, and look where the corresponding phase-space 
function~$\phi_n^\pm(p,q)$ is mainly localised (all the more sharply than $\hbar$ is small).
For a Hamiltonian
of the form~\eqref{eq:Hp2plusV} where~$V$ has local minima, some of the 
eigenstates remain localised in the neighbourhood of the stable equilibrium points. 
For instance, for a double-well potential whose shape is shown in Fig.~\ref{fig:2puits}a),
the symmetric state~$\ket{\phi_n^+}$ and the antisymmetric state~$\ket{\phi_n^-}$  at energies
below the local maximum of the barrier
have their Husimi representation localised 
 around both stable equilibrium points~$(p,q)=(0,\pm a)$, more precisely
along the lines~$H(p,q)=E$ (1d tori) at energy~$E\simeq E^+_n\simeq E^-_n$. 
Because of the non-exact degeneracy
of~$E_n^+$ and~$E_n^-$, any linear combination
of~$\ket{\phi_n^+}$ and~$\ket{\phi_n^-}$ constructed in order to be localised in 
one well only will not be stationary anymore and will oscillate back and forth 
between $-a$ and $a$ at a tunnelling frequency~$|E_n^--E_n^+|/(2\pi\hbar)$. 
At low energy  this process is classically forbidden by the energy barrier.
This is very general, even for multidimensional non-integrable systems:
the splitting $\Delta E_n=|E_n^--E_n^+|$ between some nearly degenerated doublets  
provides a quantitative manifestation of the tunnelling between the phase-space regions 
where the corresponding eigenstates are localised.

Rather than computing the poles of~$E\mapsto \tilde{G}(a,-a,E)$, 
another systematic strategy to obtain one individual
splitting \footnote{Statistical approaches have also been 
proposed~\cite{Creagh/Whelan96a,Tomsovic/Ullmo94a,Leyvraz/Ullmo96a}.}
 is to start with Herring's 
formula~\cite{Wilkinson86a,Creagh97a,Garg00a} that relies on the knowledge of the
eigenfunctions outside the classically allowed regions in phase-space. Here,
 we will propose alternative formulae \cite{Mouchet07a} that involve traces 
of a product of operators, among them the evolution operator,
\begin{multline}\label{def:U}
  \hat{U}(T)\DEF\EXP{-\imat\hat{H}T/\hbar}\\
    =\sum_{n=0}^\infty\left(
       \ket{\phi_n^+}\bra{\phi_n^+}\EXP{-\imat E_n^+ T/\hbar}+
       \ket{\phi_n^-}\bra{\phi_n^-}\EXP{-\imat E_n^- T/\hbar}\right),
\end{multline}
 analytically 
continued in some sector of the complex time domain. This approach, which privileges the time-domain, provides,
of course, a natural starting point for a semiclassical analysis in terms of
classical complex orbits.

The simplest situation occurs when the tunnelling doublet 
is made of the two lowest energies~$E_0^+\lesssim E_0^-=E_0^++\Delta E_0$ of the spectrum. 
If~$\hbar\omega$ denotes the energy difference between 
the nearest excited states and~$E_0^\pm$ (for the double well potential 
$E^\pm_1\simeq E^\pm_0+\hbar\omega$ where $\omega$ is the 
classical frequency around the stable equilibrium points), by giving a sufficiently large
imaginary part to~$-T$,
\begin{equation}\label{eq:imTlarge}
  -\omega\im(T )\gg1\;,
\end{equation} we can safely retain the $n=0$ terms only, which exponentially 
dominate  the trace of~\eqref{def:U}. To be valid, this approximation requires
that we remain away from a quantum resonance where the definition of the tunnelling doublet is made ambiguous
by the presence of a third energy level 
in the neighborhood of~$E_0^+$ and~$E_0^-$. Then 
we have immediately
\begin{equation}
  \tan\left(\frac{T\Delta E_0 }{2\hbar}\right)
 \simeq-\imat\;\frac{\tr\!\big(\opS\,\hat{U}(T)\big)}{\tr\!\big(\hat{U}(T)\big)}\;.
\end{equation}
When the tunnelling splitting is 
smaller than~$\hbar\omega$ by several orders of magnitude, we can work
with a complex time such that
\begin{equation}\label{eq:modTsmall}
  \frac{|T|\Delta E_0 }{2\hbar}\ll 1
\end{equation}
remains compatible with condition~\eqref{eq:imTlarge}  and, therefore,
\begin{equation}\label{eq:DeltaE0}
  \Delta E_0\simeq\Delta_0(T)\DEF\frac{2\hbar}{\imat T}\frac{\tr\!\big(\opS\,\hat{U}(T)\big)}
                                             {\tr\!\big(\hat{U}(T)\big)}
\end{equation}
will provide a good approximation of the tunnelling splitting for a wide range of~$T$. Though 
condition~\eqref{eq:modTsmall} is widely fulfilled in many situations this is not an essential condition
since one can keep working with $\tan^{-1}$. 
Numerically, the estimation~\eqref{eq:DeltaE0} has also the advantage of obviating a diagonalization.

When we want to compute a splitting, due to tunnelling, between an arbitrary doublet, the selection of
the corresponding terms in the right hand side of~\eqref{def:U} can be made with an operator~$\hat\varPi_n$
that will mimic the (\emph{a priori} unknown) projector~$\ket{\phi_n^+}\bra{\phi_n^+}+\ket{\phi_n^-}\bra{\phi_n^-}$.
It will be chosen such that its matrix elements are
 localised in the regions of phase-space where~$\phi_n^\pm(p,q)$ are
dominant.
Under the soft condition $|T|\Delta E_n /(2\hbar)\ll1$, we
 will therefore take
\begin{equation}\label{eq:DeltaEn}
   \Delta E_n\simeq\Delta_n(T)\DEF\frac{2\hbar}{\imat T}\frac{\tr\!\big(\opS\,\hat\varPi_n\,\hat{U}(T)\big)}
                                              {\tr\!\big(\hat\varPi_n\hat{U}(T)\big)}\;.
\end{equation}
 The localization condition on the matrix elements of~$\hat\varPi_n$
 is a selection tool that replaces~\eqref{eq:imTlarge}. In that case,
there is a battle of exponentials between the exponentially small
matrix elements~$\bra{\phi_m^\pm}\hat\varPi_n\ket{\phi_m^\pm}$
and the time dependent terms~$\EXP{-\imat(E_m-E_n)T/\hbar}$. The last term would eventually dominate
 for the lower energy states 
 ($E_m<E_n$) if~$-\im T$ could be increased arbitrarily. But once a $\im T<0$ is given,  one expects 
to recover a good approximation of the excited splitting by increasing~$\re T$ since, when~$\re T\gg|\im T|$,
we actually recover the real time case. 

 The next
 step consists in computing  $\Delta_n(T)$ by semiclassical techniques
 and, then, we will add some more specific prescriptions on the
choice of~$T$, in order to improve the accuracy of~$\Delta E_n$. 
We will illustrate how this works in the examples of secs.~\ref{sec:doublewell}, \ref{sec:pendulum}
and \ref{sec:resonanttunnelling}.

Let us mention another way to select excited states, that we did not exploit further. With the help of
a positive smooth function~$F(u)$ that has a deep, isolated minimum at~$u=0$, say $F(u)=u^{2N}$ with 
$N$ strictly positive, we can
freeze the dynamics around any energy~$E$ by considering a new Hamiltonian~$H'(p,q)\DEFt F\big(H(p,q)-E)\big)$.
Classically, the phase-space portrait is the same as the original one obtained with Hamiltonian $H$
except that the set of points~$H(p,q)=E$ now consists of equilibrium points.
The quantum Hamiltonian~$\smash{\hat{H}'\DEFt F(\hat{H}-E)}$
 has the same eigenfunctions as~$\hat{H}$ but the corresponding
 spectrum is now~$F(E_n^\pm-E)$. By choosing $E=E_n^+$, the doublet $E^\pm_n$ yields to the ground-state 
doublet~$\Delta E_n'=F(\Delta E_n)$ and we can 
use the approximation~\eqref{eq:DeltaE0} with~$\hat{U}'=\exp(-\imat\hat{H}'T)$. With the~$F$ given above, we have
\begin{equation}
   \Delta E_n\simeq[\Delta'_n(T)]^{1/(2N)}\DEF
\left[\frac{2\hbar}{\imat T}\frac{\tr\!\big(\opS\,\hat{U}'(T)\big)}
                                              {\tr\!\big(\hat{U}'(T)\big)}\right]^{1/(2N)}\;.
\end{equation}

\section{Semiclassical expressions}\label{sec:semiclassical}
\subsection{Hamiltonian dynamics with complex time}

Formally, the numerator and the denominator of the right hand side of~
\eqref{eq:DeltaE0} can be written as a phase-space path
integral of the form
\begin{equation}\label{eq:pathint}
  \int_\mathcal{P}\EXP{\imat S[p,q;t]/\hbar}\;\mathrm{D}[p]\mathrm{D}[q]
\end{equation}
where the continuous action is the functional
\begin{equation}\label{def:S}
  S[p,q;t]\DEF\int_{s_{i}}^{s_{\!f}}\left(
    p(s)\frac{\dmat q}{\dmat s}(s)-H\big(p(s),q(s)\big)\frac{\dmat t}{\dmat s}(s)\right)\dmat s.
\end{equation}
The subset~$\mathcal{P}$ of phase-space paths, the measure
$\mathrm{D}[p]\mathrm{D}[q]$ and the action~\eqref{def:S} appear as a
continuous limit ($\tau\to0$) of a discretized expression whose
precise definition depends on the choice of the basis for computing
the traces but, in any cases, involves a typical, finite, complex time
step~$\tau$ (see also appendix~\ref{app:contrib_sc}).  The complex
continuous time path~$s\mapsto t(s)$ is given with fixed ends
$t(s_i)=t_i=0$, $t(s_f)=t_f=T$. Because the slicing of~$T$ in 
small complex time steps of modulus of order~$\tau$ is arbitrary, 
the integrals of the
form~\eqref{eq:pathint} remain independent of the choice of $t(s)$
for~$s_i<s<s_f$ as long as $\im t(s)$ is non-increasing in order to
keep the evolution operators well-defined for any slice of
time~\cite{Mclaughlin72a}. In the following we will denote by~$[t]$
such an admissible time-path.  In a semiclassical limit (keeping the
order $\lim_{\hbar\to0}\lim_{\tau\to0}$
\footnote{\label{fn:ordersoflimits}The other order $\lim_{\tau\to0}\lim_{\hbar\to0}$ corresponds to the quantization
of a kicked Hamiltonian that leads to $\tau$-dependent results as shown in~\cite{Mouchet07a}.}),
the dominant contributions to integrals~\eqref{eq:pathint} 
come from some paths in~$\mathcal{P}$ that 
extremise~$S$, \textit{i.e.} from
some solutions of Hamilton's equations
\begin{subequations}\label{subeq:hamiltoneq1}
  \begin{align}
  \frac{\dmat p}{\dmat s}&=-\frac{\partial H}{\partial q}\frac{\dmat t}{\dmat s}\;;\label{eq:hamiltoneq_dpds}\\
  \frac{\dmat q}{\dmat s}&=\phantom{-}\frac{\partial H}{\partial p}\frac{\dmat t}{\dmat s}\;,
 \label{eq:hamiltoneq_dqds}
  \end{align}
\end{subequations}
with appropriate boundary conditions imposed on some canonical
variables at $s_i$ and/or $s_f$.  When $H$ is an analytic function
of the phase-space coordinates $(p,q)$, we can take the real and
imaginary parts of equations~\eqref{subeq:hamiltoneq1},  use the
Cauchy-Riemann equations that render explicit the entanglement between
the real part and the imaginary part of any analytic function~$f(z)$
: $\re(\dmat f/\dmat z)=\partial (\re f)/\partial (\re z)=\partial
(\im f)/\partial (\im z)$ and $\im(\dmat f/\dmat z)=\partial (\im
f)/\partial (\re z)=-\partial (\re f)/\partial (\im z)$, and then
obtain
\begin{subequations}\label{subeq:hamiltoneq2}
  \begin{align}
  \frac{\dmat \re p}{\dmat s}&=-\frac{\partial }{\partial \re q}\left[\re\left(H\frac{\dmat t}{\dmat s}\right)\right]\;;\\
 \frac{\dmat (-\im p)}{\dmat s}&=-\frac{\partial }{\partial \im q}\left[\re\left(H\frac{\dmat t}{\dmat s}\right)\right]\;;\\
  \frac{\dmat \re q}{\dmat s}&=\phantom{-}\frac{\partial }{\partial \re p}\left[\re\left(H\frac{\dmat t}{\dmat s}\right)\right]\;;\\
 \frac{\dmat \im q}{\dmat s}&=\phantom{-}\frac{\partial }{\partial (-\im p)}\left[\re\left(H\frac{\dmat t}{\dmat s}\right)\right]\;.
  \end{align}
\end{subequations}
Therefore, the dynamics described in terms of complex canonical variables is
equivalent to a dynamics that remains Hamiltonian --- though not autonomous with respect to the parametrisation $s$ 
whenever $\dmat t/ \dmat s$ varies with~$s$ ---,
 involving twice as
many degrees of freedom as the original system, namely $(\re q, \im q )$
and their respectively conjugated momenta $(\re p, -\im p )$. The new
Hamiltonian function is then $\re(H\dmat t/\dmat s)$ but it would have
been equivalent  (though not canonically equivalent) to choose
the other constant of motion~$\im(H\dmat t/\dmat s)$
as a Hamiltonian.

What will be important in what
follows is that $[t]$ will not be given a priori. Unlike in the
standard instanton approach where $t$ is forced to remain on the
imaginary axis, we will see that for describing tunnelling it is a
more efficient strategy to look for some complex paths~$\big(p(s),q(s)\big)$
that naturally connect two phase-space regions and then deduce~$[t]$
from one of the equations~\eqref{subeq:hamiltoneq1}. It happens that in the two usual textbook
examples that we mentioned in the first paragraph of 
our Introduction, the complex time $[t]$  has a
vanishing real part, but in more general cases where tunnelling 
between excited states is studied, this is no longer true.

\subsection{Trace formulae}

Let us privilege the $q$-representation and consider the analytic
continuation for complex time~$T$ of the well-known Van Vleck
approximation for the propagator:
\begin{equation}\label{eq:vanvleck}
  G(q_f,q_i;T)\scl\sum_{\goo}(-1)^{\kappa_\goo}
                 \sqrt{\det\left(\frac{\imat}{2\pi\hbar}\frac{\partial^2 S_\goo}{\partial q_i\partial q_f}
                            \right)
                      }\;
                   \EXP{\imat S_\goo/\hbar}\;.
\end{equation}
The sum involves (complex) classical trajectories~$\goo$ in
phase-space \textit{i.e.}  solutions of equations~\eqref{subeq:hamiltoneq2}
with~$q(s_i)=q_i$, $q(s_f)=q_f$ for a given~$[t]$ such
that~$t(s_i)=0$, $t(s_f)=T$. The action $S_\goo$ is computed
along~$\goo$ with definition~\eqref{def:S} and is considered as a function
of~$(q_f,q_i;T)$. The integer~$\kappa_\goo$ encapsulates the choice of the Riemann sheet
where the square root is computed; it keeps a record of the number of points on~$\goo$ where
the semiclassical approximation~\eqref{eq:vanvleck} fails.
 As far as we do not cross a bifurcation 
of classical trajectories when smoothly deforming~$[t]$, the number of~$\goo$'s, the value of
$S_\goo$ and~$\kappa_\goo$ do not depend of the choice of~$[t]$.
  The numerator
(resp. denominator) of~$\Delta_0(T)$ are
given by the integral~$\int G(\eta q,q;T)\dmat q$ where $\eta=-1$
(resp. $\eta=+1$).
Within the semiclassical approximation, when~$\goo$ contributes 
to the propagator~$G(\eta q,q;T)$, we have to evaluate
\begin{equation}\label{eq:intq}\
 (-1)^{\kappa_\goo}\!\!\!
  \int\dmat q \sqrt{\det\left(\frac{\imat}{2\pi\hbar}
                               \frac{\partial^2 S_\goo}{\partial q_i\partial q_f}\bigg|_{(\eta q, q; T)}\right)}\;
                   \EXP{\imat S_\goo(\eta q,q,T)/\hbar}\;.
\end{equation} 
The
steepest descent method requires the determination of the
critical points of~$S_\goo(\eta q, q, T)$. Since the momenta at the
end points of $\goo$ are given by
\begin{subequations}\label{subeq:pipfS}
  \begin{eqnarray}
  p_i=&p(s_i)&=-\partial_{q_i}S_\goo(q_f,q_i;T)\;;\\
  p_f=&p(s_f)&=\phantom{+}\partial_{q_f}S_\goo(q_f,q_i;T)\label{eq:pf}\;, 
 \end{eqnarray}
\end{subequations}
the dominant contributions to $\tr\!\big(\hat{U}(T)\big)$ come from periodic 
orbits, \textit{i.e.} when $(p_f,q_f)=(p_i,q_i)$,
whereas  the dominant contributions 
to  $\tr\!\big(\opS\hat{U}(T)\big)$ come from half \emph{symmetric} periodic orbits,
\textit{i.e.} when~$(p_f,q_f)=(-p_i,-q_i)$.
As explained in detail in the appendix~\ref{app:contrib_sc}, one must distinguish  the
contributions of the 
zero length orbits~$\goe$ (the equilibrium points)
from the non-zero length periodic orbits~$\goo$. For one degree of freedom, 
their respective contributions 
are given, up to sign, by
\begin{subequations}\label{subeq:contrib_gene}
\begin{equation}\label{eq:contrib_e}
  \frac{\EXP{-\imat H(p_e,q_e)T/\hbar}}
       {\EXP{\lambda_\goe T/2}-\eta\EXP{-\lambda_\goe T/2}}\;,
\end{equation}
with~$\pm\lambda_\goe$ the two Lyapunov exponents of the equilibrium point~$\goe=(p_e,q_e)$, 
and 
\begin{equation}\label{eq:contrib_ppo}
   (-1)^{\mu_\goo}
  \frac{\sum_\beta T_{\beta}}{\sqrt{-2\eta\imat\pi\hbar}}
  \sqrt{\frac{\dmat E_\goo}{\dmat T}}\;\EXP{\imat S_\goo/\hbar}\;,
\end{equation}
\end{subequations}
where $\goo$ is a periodic orbit of period~$T_\goo=T$ if $\eta=+1$ and  
half a symmetric periodic orbit of
half period~$T_\goo=T$ if $\eta=-1$. The sum runs over all the 
branches~$\beta$ that 
compose the geometrical set of points belonging to~$\goo$. 
$T_{\beta}$ is the characteristic 
time~\eqref{def:primitiveperiod} on the branch~$\beta$. The energy $E_\goo$ is 
implicitly defined by~\eqref{eq:TEqiqf} and~$\mu_\goo$ essentially 
counts the number of turning points on~$\goo$.

Expressions~\eqref{subeq:contrib_gene} are purely geometric;  their classical ingredients 
do not rely on a specific choice of canonical coordinates and 
they are independent of the choice of
the basis to evaluate the traces.

In the general case, the most difficult part consists in determining which 
periodic orbits contribute to the semiclassical approximation of the traces. 
Condition~$T_\goo=T$ is necessary but far from being sufficient;  the structure of the complex
paths (keeping  a real time) may appear to be very subtle \cite{Shudo/Ikeda96a} and 
have been the subject of many recent delicate works \cite{Shudo+09a}.
Even for a simple oscillating integral, the determination of 
the complex critical points of the phase that do contribute is a highly non-trivial 
problem because it requires a global analysis: one must know how to deform the whole initial contour of 
integration to reach a steepest descent paths.

Our strategy consists in retaining the terms \eqref{eq:contrib_ppo} for which we can 
choose the complex time~$[t]$  to constrain the (half) periodic orbit~$\goo$ to keep
one of the canonical coordinate (say $q$) real.
This~$\goo$  surely  contributes because 
 we do not have to deform the $q$ part of the integration 
domain of~\eqref{eq:pathint} in the complex plane. But we will see in the next section 
that two different
periodic orbits with real~$q$ correspond to two different choices of~$[t]$: 
For a chosen~$[t]$, only some isolated points, if any, in phase space, will provide 
a starting point~$(p_i,q_i)$ of a trajectory with $q(s)$ real all along~$\goo$. 
When sliding slightly the initial real coordinate~$q_i$ in the integral~\eqref{eq:intq},  it requires 
to change~$[t]$ as well to maintain~$q(s)$ real on the whole~$\goo$. This is
not a problem since all the quantities involved in expression~\eqref{eq:intq} 
are~$[t]$-independent if the deformation
of~$[t]$ is small enough not to provoke a bifurcation of~$\goo$, that is, whenever the initial
point does dot cross one of the turning points, which are the boundaries of
 the branches~$\beta$.
The integral~\eqref{eq:intq} may be computed with a fixed time-path or an adaptive one for each 
separated branch. This computation, presented in the appendix~\ref{app:contrib_sc},
generates the sum over all branches that appears in formula~\eqref{eq:contrib_ppo}. 
To sum up, to keep a contribution to the traces, we must  know if we can choose a shape of~$[t]$ in
order to pick a real-$q$ periodic orbit with a specific~$q_i$. 

 This construction is certainly 
not unique (one may choose other constraints) so  we will use the intuitive principle that the periodic orbits we choose
will connect the two regions of phase-space that are concerned by tunnelling \textit{i.e.}
where~$\phi_n^\pm$ are dominant; this is justified by the presence of the operator~$\hat\varPi_n$
in the right hand side of~\eqref{eq:DeltaEn}. The prefactor~$\varPi_n(q,q')\DEF\bra{q}\hat\varPi_n\ket{q'}$
will select in the integral
\begin{equation}\label{eq:intqqprimePin}
  \int\dmat q\,\dmat q' \varPi_n(q,q')\, G(\eta q',q;T)\;,
\end{equation}  
a domain around the projection onto the~$q$-space of the appropriate tori. 
 In the specific case of the ground-state splitting, condition~\eqref{eq:imTlarge}  
does the job of $\hat\varPi_0$: the large value of~$\im T$  requires that
 the orbit approach at least one equilibrium point and then follow a separatrix line.
 In general we will not be able to prove that  other complex periodic orbits
give sub-dominant contributions but the examples given in the following sections are rather convincing.
Moreover, our criterion of selection allows us to justify the four rules presented 
in~\cite[sec.~IIA]{Robbins+89a} for computing  the contributions of
complex orbits to the semiclassical expansions of the energy Green's function.

\section{Application to the double-well potential}\label{sec:doublewell}

\begin{figure}[!ht]
\center
\includegraphics[width=8.5cm]{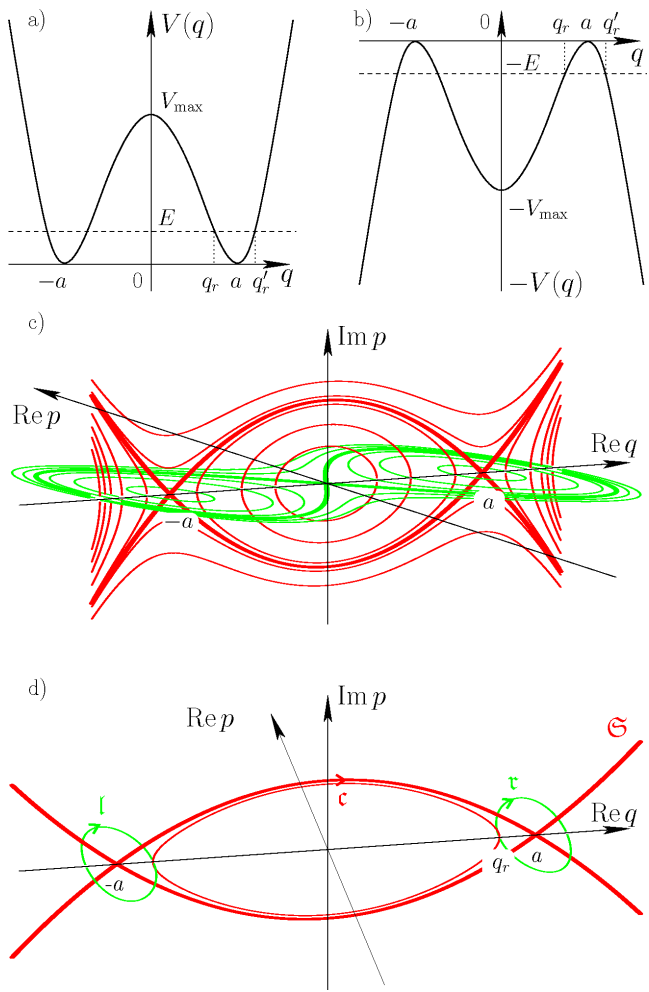}
\caption{\label{fig:2puits} (colour online) In the case of a double
  well potential shown in a), tunnelling can be described by a
  solution of~\eqref{subeq:hamiltoneq1} with a purely real $q$
  evolving from one well to the other. In phase-space, this trajectory
  appears to be a concatenation of two types of curves that join on
  the $(\re q)$-axis: 1)~a trajectory that lies in the phase-space
  plane~$(\re q,\re p)$ with real variations of $t$ at energy~$E$ with the
  potential~$V$ and 2)~a trajectory that lies in the phase-space
  plane~$(\re q,\im p)$ with imaginary variations of $t$ at energy~$-E$ with the
  potential~$-V$ shown in b).  In c) a family of constant energy curves is
  shown (horizontal green (light gray) for the first type, vertical 
red (dark gray) for the
  second type).  In d) for a given energy, we show how the three
  curves~$\gol,\goc,\gor$ glue together at the turning points~$(0,\pm
  q_r)$.  $\goS$ denotes the separatrix $\im p=\pm\sqrt{2V(q)}$.}
\end{figure}

Let us show first that our strategy leads to the usual instanton results for a Hamiltonian of
the form~\eqref{eq:Hp2plusV} with an even~$V$ having two stable symmetric equilibrium points at~$q=\pm a$
(Fig.~\ref{fig:2puits}a);  in their neighbourhood, the frequency of the small vibrations is~$\omega$.
We are interested in the ground-state splitting~$\Delta E_0$ for~$\hbar$ small enough in order to 
have~$\Delta E_0\ll\hbar\omega$.

\begin{figure}[!ht]
\center
\includegraphics[width=8.5cm]{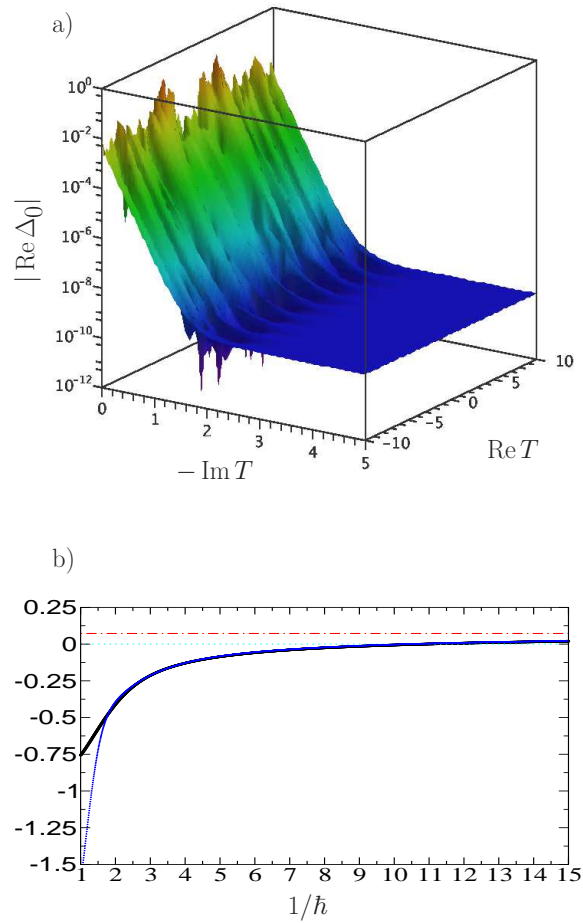}
\caption{\label{fig:Delta0_2puits} (colour online) For~$V(q)=(q^2-a^2)^2$ 
with~$a=1$ and~$\hbar\simeq1/12$, we have plotted
in the upper graph a) the real part of~$\Delta_0$ defined by~\eqref{eq:DeltaE0} as a function of the complex~$T$.
It becomes constant  for $\im T$ large enough for having \eqref{eq:imTlarge} ($1/\omega=1/(2\sqrt{2}a)\simeq.35$) 
regardless of~$\re T$. In the same range, $\im\big(\Delta_0(T))$ is
negligible compared to $\Delta E_0\sim 4.4\;10^{-10}$ and condition~\eqref{eq:modTsmall} is largely fulfilled.
In the lower graph~b), we have computed the
``exact" value~$\Delta E_0$ by direct diagonalization. The black thick 
solid line
corresponds to~$\ln(\Delta E_0)$ from which we have substracted 
$\Lambda(\hbar)\DEFt-\tilde{S}_\goc(\hbar\omega/2)/(2\hbar)+\ln(\hbar\omega/\pi)$
in order to emphasize the contribution of the prefactor.
The blue solid line corresponds to~$\ln|\Delta_0|-\Lambda(\hbar)$ for~$T\simeq-4\imat$ and 
become indistinguishable from the previous one
for~$1/\hbar\gtrsim 3$. With  the same substraction, our first proposal~\eqref{eq:Delta_0Garg}
 gives the constant~$(\ln\sqrt{\pi})\simeq0.6$  and our second one~\eqref{eq:Delta_nGarg},
which is the same as the formula in Landau and Lifshitz~\cite[sec.~50, problem~3]{Landau/Lifshitz77a}, corresponds to~$0$ (dotted line).
 Garg's formula~\cite[eq.1.1]{Garg00a} gives the constant~$(\ln\sqrt{\pi/\mathrm{e}}\simeq0.07)$ (dash-dotted line).}
\end{figure}

 Before 
we make any semiclassical approximations, we check in Fig.~\ref{fig:Delta0_2puits} the validity of 
the estimation~\eqref{eq:DeltaE0} on the quartic potential
 by a) verifying that $\Delta_0(T)$ is almost real and independent 
of the choice of the complex~$T$
provided that conditions~\eqref{eq:imTlarge} and~\eqref{eq:modTsmall} are fulfilled,
and b) by checking that this constant gives a good approximation of the ``exact"~$\Delta E_0$
computed by direct numerical diagonalization of the Hamiltonian~\cite{Korsch/Gluck02a}.

\begin{figure}[!ht]
\center
\includegraphics[width=17cm]{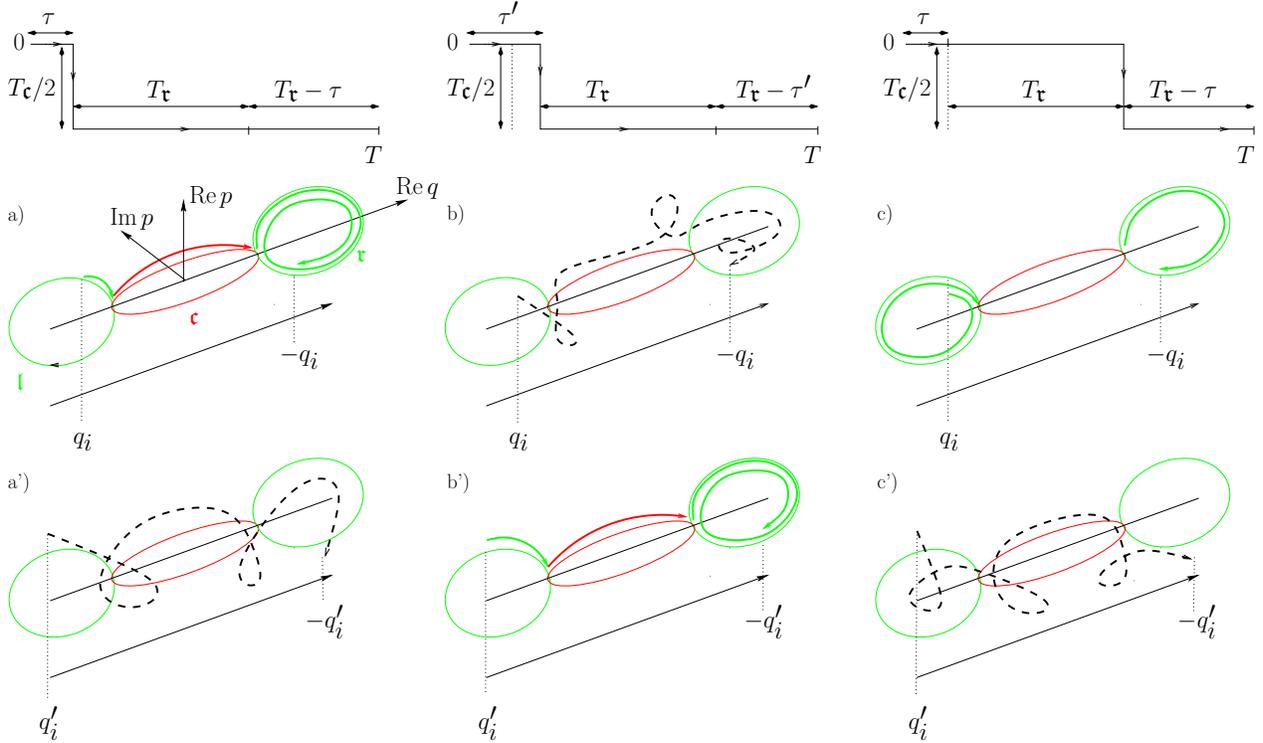}
\caption{\label{fig:slidingstaircase} (colour online) For the double well in Fig.~\ref{fig:2puits}, at a given energy~$0<E<V_{\mathrm{max}}$, 
when~$T=2T_\gor(E)-\imat T_\goc(E)/2$, for each one-step 
time path, there exists a unique half  symmetric
periodic orbit with energy~$E$ keeping~$\im q=0$ starting from the left allowed region. Its initial point is uniquely defined:
 For the time-path depicted in the left column, $(p_i,q_i)$ is such that it takes exactly the real time~$\tau$ to reach
the turning point~$(0,-q_r)$ following the primitive orbit~$\gol$. 
Then it follows   half of the orbit~$\goc$ and ends in the right well winding once along~$\gor$ (a).
 When starting at~$q_i'<q_i$, there still exists a half periodic orbit joining~$(p'_i,q'_i)$ to~$(-p'_i,-q'_i)$
in time~$T$ for some~$p'_i$ but unlike the previous one it gets outside the three dimensional section~$\im q=0$ (its projection is
schematically represented by the dashed line in a'). However an appropriate shift of the time step to~$\tau'>\tau$ (central column) 
turns  the latter smoothly into a trajectory with~$\im q=0$ (b').  During this process, the first trajectory starting at~$q_i$
cannot keep a purely real~$q$ anymore (b). If we shift the step by exactly~$T_\gor$ (right column), we recover a trajectory 
starting at~$q_i$ with purely real~$q$ (c) but with a different topology since it is now winding once along~$\gol$. The change of topology occurs when
the initial position reaches a turning point while sliding the step.}
\end{figure}

As explained at the end of section~sec.~\ref{sec:semiclassical}, in 
phase-space we will try to find  some time-path~$[t]$ that allows the existence a (half) symmetric 
periodic orbit~$\goo$ that connects 
two tori at (real) energy~$E\gtrsim0$ in the neighbourhood of~$(p,q)=(0,\pm a)$ 
while $q$ remains real. If we impose~$t(s_i)=0$ and~$t(s_f)=T$, then $E=E_\goo(q_f,q_i,T)$ is
 implicitly given by relation~\eqref{eq:TEqiqf}. We will denote by~$q_r(E)$ (resp. $q'_r(E)$)
the position of the turning  point at energy~$E>0$ that lies in between $q=0$ and~$q=a$ 
(resp. that is larger that~$a$). The two branches $p_\pm(q,E)=\pm\sqrt{2\smash{\big(}E-V(q)\smash{\big)}}$ are either purely 
real when~$V(q)\geqslant E$ or
purely imaginary when~$V(q)\leqslant E$. Then, from equation~\eqref{eq:hamiltoneq_dqds},
\begin{equation}
  \frac{\dmat t}{\dmat s}=\frac{1}{p}\,\frac{\dmat q}{\dmat s}
\end{equation}
is purely real in the classically allowed region, while purely imaginary in the forbidden region. 
Therefore the complex time path $[t]$ must have the shape of a descending staircase whose steps 
are made of pure real or pure imaginary variations of time (see Fig.~\ref{fig:slidingstaircase}). 

 The
 complex  orbit with real~$q$  can be represented in  phase-space 
as a continuous concatenation of paths, following the lines~$E=(\re p)^2/2+V\big(q)$ in the allowed region
and $-E=(\im p)^2/2-V(q)$ in the forbidden region. 
 It is natural to represent~$\goo$ in the three dimensional section~$\im q=0$ of the complex phase-space
with axes given by~$(\re q,\re p,\im p)$: the junctions at the turning points lie necessarily on 
the~$(\re p=0,\im p=0)$ axis (see Figs.~\ref{fig:2puits}c and~d). 
A periodic orbit 
 is made of a succession
of repetitions of \\
(i) primitive real periodic orbits~$\gor$ with energy~$E$ in the right region,
 that is, such that $q_r(E)\leqslant q(s)\leqslant q_r'(E)$ and $\im p(s)=0$;\\
(ii) primitive complex periodic orbits  
$\goc$ in the central region with purely imaginary~$p$ and real $q$,
 that is, such that~$-q_r(E)\leqslant q(s)\leqslant q_r(E)$ and $\re p(s)=0$;\\
(iii) primitive real periodic orbits~$\gol$ with energy~$E$ in the left region. They are obtained
from the periodic orbits~$\gor$ by the symmetry~$\mathsf{S}$.

 By denoting~$T_\gor(E)$ and~$T_\goc(E)$ 
the (real, positive) periods of the primitive periodic orbits~$\gor$ and $\goc$ at energy~$E$ respectively, we have 
\begin{equation}\label{eq:Tw1w2}
  T_\goo(E)=w_\gor T_\gor(E)-\imat w_\goc T_\goc(E)-\imat\frac{1-\eta}{4}\, T_\goc(E)
\end{equation}
with the winding numbers~$w_\gor$ and~$w_\goc$ being non-negative integers. 
 The trajectory~$\goo$ may contribute to the denominator ($\eta=+1$) or
 to the numerator ($\eta=-1$) of the right hand side of~\eqref{eq:DeltaE0}
provided that~$T_\goo=T$.   For $\eta=+1$,
 $\goo$ is periodic whereas, for $\eta=-1$, $\goo$ is half a symmetric periodic orbit.
We will keep the contributions of all orbits for which a staircase $[t]$ can be constructed.
We can understand from figure~\ref{fig:slidingstaircase} that
orbits differing by a small sliding of their initial~$q_i$ can be obtained
by a small modification of the length or height of the first step. All these contributions are summed up 
 when performing the integral~\eqref{eq:intq} on one branch~$\beta$ 
and correspond to one term in the sum~$\sum_\beta$ in formula~\eqref{eq:contrib_ppo}.
 When at least one of the winding number is strictly larger than one,
several staircase time-paths can be constructed 
while keeping relation~\eqref{eq:Tw1w2}: they differ one from each other
by a different partition into steps of the length and/or heights of the staircase. The corresponding 
orbits~$\goo$
can be obtained one from each other by a continuous smooth sliding of the steps of the staircase but 
during this process one cannot avoid an orbit starting at a turning point where a bifurcation occurs.

\onecolumngrid

\twocolumngrid
In the right hand side of~\eqref{eq:vanvleck},
 the  sum involves several trajectories differing 
one from each other by the sequence of turning points that are successively encountered along~$\goo$.
Therefore, to compute the dominant contribution of the non-zero length orbits
to the numerator and to the denominator 
of~\eqref{eq:DeltaE0}, we
will add all the contributions
of the topological classes of orbits, each of them uniquely 
characterised by an ordered sequence 
of turning points~$[\rho_1,\rho_2,\dots]$, in other words by a partition of~$w_\gor$ and~$w_\goc$
into integers and by the branch~$\beta$ where its starting point lies. 
We can therefore express
our result in a way that can be applied to cases more general than the double well:  the total
contribution of the non-zero length paths to the numerator ($\eta=-1$) and to the denominator 
($\eta=+1$) of~\eqref{eq:DeltaE0} is
\begin{equation}\label{eq:generalsum}
  \sum_{\gos}\sum_{[\rho_1,\rho_2,\dots]}(-1)^{\mu_\goo}\frac{T_{\beta}}{\sqrt{-2\eta\imat\pi\hbar}}
  \sqrt{\frac{\dmat E_\goo}{\dmat T}}\;\EXP{\imat S_\goo/\hbar}\;.
\end{equation}
$\gos$ denotes a section of an energy surface in complex phase space corresponding
to one purely real canonical variable. 
$[\rho_1,\rho_2,\dots]$ is an ordered sequence of (not necessarily distinct) turning points  that belong to
a section~$\gos$. The sum concerns all~$\gos$ (different energies may 
be possible) and $[\rho_1,\rho_2,\dots]$
 such that we can construct on~$\gos$, with an appropriate choice of~$[t]$, a
 periodic orbit~$\goo$  if $\eta=+1$  (a half symmetric periodic orbit~$\goo$ if $\eta=-1$)
of period $T_\goo=T$. The branch~$\beta$ is the one where~$\goo$ starts, 
the sequence of turning points that are successively 
crossed by~$\goo$ is exactly $[\rho_1,\rho_2,\dots]$.

In the case of the double-well, for an energy below~$V_\mathrm{max}$, a section~$\gos$ for real~$q$
has four turning points~$(0,\pm q_r)$ and~$(0,\pm q'_r)$. Only the points on~$\gos$
such that~$-q'_r<q<q'_r$ can provide starting points of a periodic orbit.  They belong 
to one of the three closed loops~$\gor$,$\goc$, $\gol$ that connect on the axis~$(p=0, \im q=0)$
at the turning points~$(0,\pm q_r)$. Once~$T$ is given, the condition~$T_\goo=T$
will select a finite set of energies for 
which we have \eqref{eq:Tw1w2} with integers~$w_\gor$ and~$w_\goc$.
 Any~$\goo$ with winding 
numbers~$w_\gor$ and $w_\goc$ will have an action given by
\begin{equation}\label{eq:Sow1w2}
  S_\goo=w_\gor S_\gor+\imat w_\goc S_\goc+\imat\frac{1-\eta}{4}\, S_\goc
\end{equation}
and an index 
\begin{equation}\label{eq:muow1w2}
  \mu_\goo=w_\gor+w_\goc+\frac{1-\eta}{2}
\end{equation}
(in the real case, $2\mu_\goo$ is computed in the same way as the Maslov index: it counts the
number of turning points encountered along~$\goo$).
These quantities are independent on the choice of the six possible starting branch~$\beta$
(each~$\gor$, $\goc$, $\gol$ is made of two branches). When the orbit~$\goo$ starts 
on~$\gor$ or on~$\gol$, we will have~$T_\beta=T_\gor/2$ and when~$\goo$ starts 
on~$\goc$, we have~$T_\beta=-\imat T_\goc/2$. 
 
For $-\im T$ larger than the oscillation period in the central well of~$-V$, $w_\goc=1$
 is the minimum value of the winding number when~$\eta=+1$ (just one back and forth trip around~$\goc$)
while it is~$w_\goc=0$ when $\eta=-1$ (just one half of~$\goc$ is concerned). 
For these orbits, condition~\eqref{eq:imTlarge} forces $\goc$ to stay
 near the separatrix~$\goS$ defined by~$\im p=\pm\sqrt{2V(q)}$ and, 
thus, $\gor$ must lie in the immediate neighbourhood
of the equilibrium point~$(p,q)=(0,a)$. These orbits will give the dominant contribution because
they have the smallest $\im S_\goo$ among all the other possible orbits involving repetitions of $\goc$.
Indeed, for a fixed $T$, all the orbits~$\goo$ that may contribute semiclassically are such 
that~$\im S_\goo=-2\im(T)S_\goc/T_\goc$ 
and  $S_\goc/T_\goc$ is a decreasing function of the energy~$\tilde E\DEFt-E$ 
when~$\goc$ is inside the separatrix since
\begin{multline}
  \frac{\dmat}{\dmat \tilde E}\left(\frac{S_\goc(\tilde E)}{T_\goc(\tilde E)}\right)=\\
  \frac{\dmat}{\dmat \tilde E}\left(\frac{4}{T_\goc(\tilde E)}\int_{0}^{q_r(-\tilde E)} \sqrt{2\big(\tilde E+V(q)\big)}\dmat q-\tilde E\right)
   \\=-\frac{4}{T_\goc(\tilde E)^2}\frac{\dmat T_\goc(\tilde E)}{\dmat \tilde E}\int_{0}^{q_r(E)} |p(q,E)|\dmat q <0\;.
\end{multline}
Therefore $\im S_\goo$ reaches its minimum when $\tilde E$ is at its maximum, that is~$E\to0^+$. The only
possible equilibrium point contributing 
to $\tr\!\big(\opS\,\hat{U}(T)\big)$ is 
the origin~$(p_f,q_f)=(-p_i,q_i)=(0,0)$; it is also  subdominant because~$H(0,0)=V(0)=V_{\textrm{max}}$ 
 (which  can be seen as the limit of $S_\goc(\tilde E)/T_\goc(\tilde E)$ when~$\tilde E\to-V_{\textrm{max}}$)
is strictly  larger than $S_\goc(\tilde E)/T_\goc(\tilde E)$ for~$\tilde E>-V_{\textrm{max}}$.

Assuming that only
orbits with real~$q$ do contribute to~$\tr\!\big(\opS\,\hat{U}(T)\big)$, we have proven that 
 the dominant contribution is given by the half symmetric orbits~$\goo$ at energy $E$ such that
$T_\goc(E)=-2\im T$ ($w_\goc=0$). However, for such an orbit to exist we cannot choose~$\re T$ arbitrarily
since it must be an integer multiple of~$T_\gor(E)$. To put it in another way, for a given~$E$, 
we will choose~$T=w_\gor T_\gor(E)-\imat T_\goc(E)/2$ such that an orbit~$\goo$  with a real $q$ exists.
Condition~\eqref{eq:imTlarge} will be fulfilled if~$E$ is sufficiently small. 
We must now enumerate all the topological classes concerned by the sum~\eqref{eq:generalsum}:
When~$\goo$ starts on the upper branch~$\im p>0$ of~$\goc$, it will reach the turning point~$\rho_1=(0,q_r)$
then wind~$w_\gor$ times around~$\gor$ alternatively crossing~$(0,q_r')$ and~$(0,q_r)$
 before turning back on the lower branch~$\im p<0$ of~$\goc$. Starting on the lower branch 
 of~$\goc$ corresponds to the symmetric trajectory and provide the same contribution 
with~$T_\beta=-\imat T_\goc/2$.
When starting on the upper branch of~$\gor$, $\goo$  crosses~$\rho_1=(0,q'_r)$ first, then 
reaches~$\rho_2=(0,q_r)$. Then it may go on  winding~$r$ times around~$\gor$ then take the lower 
branch of~$\goc$ up to the turning point $(0,-q_r)$, wind~$w_\gor-1-r$ times around~$\gol$ and eventually
join the symmetric of its starting point on the lower branch of~$\gol$. There are exactly~$w_\gor$ 
such topological classes because we can take~$r=0,...,w_\gor-1$. If we start from the lower branch of~$\gor$
or on one of the two branches of~$\gol$, we obtain the same contribution and exhaust the possible topological classes. 
The sum of $T_\beta$ on all~$\beta$'s and classes is then~$2 (-\imat T_\goc/2)+4w_\gor T_\gor/2=2T$ and keeping only
the $w_\goc=0$ solutions, the sum~\eqref{eq:generalsum} reduces to
\begin{equation}
  \tr\!\big(\opS\,\hat{U}(T)\big)\scl
   \frac{2 T(-1)^{w_\gor+1}}{\sqrt{2\imat\pi\hbar}}\sqrt{\frac{\dmat E_\goo}{\dmat T}}\;
    \BIGEXP{ \imat S_\goo/\hbar}
\end{equation}
with~$\goo$ being one half symmetric orbit of energy~$E_\goo$  defined implicitly by
\begin{equation}\label{eq:TTrTc}
  T=w_\gor T_\gor(E)-\imat T_\goc(E)/2
\end{equation}
 with
\begin{subequations}
\begin{align}
   T_\gor(E)&=2\int_{q_r(E)}^{q'_r(E)}\frac{\dmat q}{\sqrt{2\big(E-V(q)\big)}}\;,\\
   T_\goc(E)&=4\int_0^{q_r(E)}\frac{\dmat q}{\sqrt{2\big(V(q)-E\big)}}\;.\label{eq:Tc}
\end{align}
\end{subequations}
We have~$S_\goo(E)=w_\gor S_\gor(E)+\imat S_\goc(E)/2$ with
\begin{subequations}\label{def:Stilde}
\begin{align}
  S_\gor(E)+ET_\gor(E)&=\tilde{S}_\gor(E)\DEF2\int_{q_r(E)}^{q'_r(E)}\!\!\sqrt{2\big(E-V(q)\big)}\,\dmat q, \\ 
\vspace{-2\baselineskip}
 S_\goc(E)- ET_\goc(E)&=\tilde{S}_\goc(E)\DEF4\int_0^{q_r(E)}\!\!\sqrt{2\big(V(q)-E\big)}\,\dmat q.
\end{align}
\end{subequations}
The dominant contributions to~$\tr\!\big(\hat{U}(T)\big)$ comes from the
 two stable equilibrium points~$\goe=(0,\pm a)$
for which~$\lambda_\goe=\imat\omega$.
The contribution of $(0,0)$ is sub-dominant as well as the contribution of 
any periodic orbit~$\goo$ which necessarily
turns around about~$\goe$ during~$\re T$  then follow an orbit~$\goc$  near~$\goS$ during~$-\im T$ before coming back to 
its initial point. Together with
\eqref{eq:imTlarge}, $\eta=1$, the two stable equilibrium points give two identical contributions
\eqref{eq:contrib_e} and we have
\begin{equation}
  \tr\!\big(\hat{U}(T)\big)\scl2\,\EXP{-\imat \omega T/2}
\end{equation}
which of course could have been deduced directly from~$E_0^\pm\simeq\hbar\omega/2$.
Collecting all these results in the right hand side of~\eqref{eq:DeltaE0}, 
we obtain
\begin{multline}\label{eq:Delta0sc}
  \Delta_0\scl\sqrt{\frac{2\imat\hbar}{\pi}}\left(\frac{\dmat T_\goo}{\dmat E}\right)^{-1/2}\\
 \times \exp{\left(-\frac{1}{2\hbar}\Big(\tilde{S}_\goc(E)+(E-\hbar\omega/2)T_\goc(E)\Big)\right)}\\
 \times\,\exp{\left(\frac{\imat w_\gor}{\hbar}\Big(\tilde{S}_\gor(E)-(E-\hbar\omega/2) T_\gor(E)-\hbar\pi\Big)\right)}.
\end{multline}

When~$E\to0^+$, we have the following asymptotic expansions  (see appendix~\ref{app:Sasymptotic})
\begin{subequations}\label{subeq:Sasymptotique}
\begin{align}
  \tilde{S}_\goc(E)&=\tilde{S}_\goc(0)+\frac{4E}{\omega}\ln\bigg(\frac{\sqrt{2E}}{2a\omega}\bigg)
  -\frac{2(2A+1)}{\omega}E+\mathrm{o}(E)\;;\label{eq:Scasymptotique}\\
 \tilde{S}_\gor(E)&=\frac{2\pi E}{\omega}+BE^2+\mathrm{o}(E^2)\;;\label{eq:Srasymptotique}
\end{align}
\end{subequations}
with
\begin{align}
  A&\DEF\int_0^a\left(\frac{\omega}{\sqrt{2V(q)}}-\frac{1}{a-q}\right)\,\dmat q\;;\label{def:A}\\
  B&\DEF\frac{\pi}{24\omega^7}\big(5V^{(3)}(a)^2-3\omega^2V^{(4)}(a)\big)\label{def:B}
\end{align}
(the superscript in parenthesis indicates the order of the derivative
of~$V$).  The differentiation of
expressions~\eqref{subeq:Sasymptotique} with respect to~$E$ leads to
the asymptotic expansions for~$-T_\goc$ and~$T_\gor$. From the first
one we can extract the exponential sensitivity of~$E$ on~$\im T$:
\begin{equation}\label{eq:ET}
  E\simeq 2a^2\omega^2\EXP{2A}\EXP{\omega\im T}\;.
\end{equation}
From relation~\eqref{eq:TTrTc} we can see 
that $2\dmat T_\goo/\dmat E
=2w_\gor\,\dmat T_\gor/\dmat E-\imat\,\dmat T_\goc/\dmat E$ is dominated by the last term if~$E$ is small.
\begin{equation}\label{eq:}
 \frac{\dmat T_\goo}{\dmat E}
\simeq-\frac{\imat}{2}\,\frac{\dmat T_\goc}{\dmat E}
     \simeq\frac{\imat}{\omega E}\;.
\end{equation}
Inserting all these asymptotic expansions  in the right hand side of~\eqref{eq:Delta0sc}, we get
\begin{multline}
   \Delta_0\scl\sqrt{\frac{4\hbar a^2\omega^3}{\pi}}\EXP{A}\EXP{E/(\hbar\omega)}\,\EXP{- \tilde{S}_\goc(0)/(2\hbar)} \\
            \times \EXP{-\imat w_\gor BE(E-\hbar\omega)/\hbar}\;.
\end{multline}
This expression can be turned into the usual \textsc{jwkb} expansion~$\exp\big(a_0(E)/\hbar+a_1(E)\ln\hbar+a_2(E)+\mathrm{o}(1)\big)$: 
As soon as we have condition~\eqref{eq:imTlarge}, from expression~\eqref{eq:ET}
we see that~$E$ is exponentially small
and we obtain~$a_0(0)=-\tilde{S}_\goc(0)/2$ and $a_1(0)=1/2$. To obtain the correct value of~$a_2$, 
we must proceed to  fine-tune the choice of~$T$. A criterion is to impose on~$\Delta_0$
to have a vanishing  imaginary part at any order in~$\hbar$ consistent with the \textsc{jwkb}
expansions used so far. From~\eqref{eq:Delta0sc}, we will choose $E$ such that~$\tilde{S}_\gor(E)=\hbar\pi$,
which is exactly the Einstein-Brillouin-Keller quantization condition for the ground state in one well. This leads
to~$E=\hbar\omega/2+\mathrm{o}(\hbar)$. Then~$T_\goc=-2\ln\big(\hbar/(4a^2\omega)\big)/\omega+4A/\omega+\mathrm{o}(1)$ and
\begin{equation}\label{eq:Delta_0Garg}
  \Delta_0\scl\frac{\hbar\omega}{\sqrt{\pi}}\EXP{- \tilde{S}_\goc(\hbar\omega/2)/(2\hbar)}
 =2a\omega\sqrt{\frac{\mathrm{e}\hbar\omega}{\pi}}\EXP{A}\EXP{- \tilde{S}_\goc(0)/(2\hbar)},
\end{equation}
which differs from~\cite[eqs.~(1.1) and~(1.4)]{Garg00a} by a reasonable factor~$\sqrt{\mathrm{e}}\simeq1.6$.
This discrepancy, already noticed in~\cite[sec.~V]{Gildener/Patrascioiu77a}, which appears in the third order term in the $\hbar$-expansion, comes
 from the different kind of approximations involved in our approach on the one hand 
and in Herring's formula on the other hand. 

We are also able to obtain a formula for the splitting of the excited
states that is consistent with the result given
in~\cite[eq.~(B1)]{Garg00a}. Using a semiclassical approximation for
the matrix element of~$\hat\varPi_n$, we explain in detail in
appendix~\ref{app:contrib_sc_Pin} how to obtain $\Delta_n(T)$.  For
one dimensional systems whose energy surface~$E_n$ is made of two
branches (two Riemann sheets in the complex plane), we can
insert~\eqref{eq:Abetabetap2branches} into~\eqref{eq:scDeltangeneral}
and get one of the main result of this paper:
\begin{equation}\label{eq:scDeltan}
  \Delta_n(T)\scl\frac{\hbar}{2T}\sum_{[\rho_1,\rho_2,\dots]}(-1)^{\mu_\goo+1}
\EXP{\imat\tilde{S}_\goo(b_{\beta'},b_{\beta^{}},E_n)/\hbar}\;.
\end{equation} 
To see how formula \eqref{eq:scDeltan} works in the case of the double-well potential,
we choose the quasi-mode~$\ket{\Phi_n}$ localised on the right torus~$\gor$ at energy~$E_n$.
This torus is made of two branches labelled by the sign of~$p$
and we can choose a common base point for these two branches, namely
$b_\pm=q_r(E)$.  On the symmetric torus~$\gol$, the two base points will be~$b_\pm=-q_r(E)$.
 Then $\tilde{S}_\goo(b_{\beta'}, b_{\beta^{}},E_n)$ and the index~$\mu_\goo$ do not depend on the choice 
of the initial and final branch. The orbits~$\goo$  that go from~$\gor$ to~$\gol$
must correspond to a~$T$ of the form
\begin{equation}
  T=\tau + w_\gor T_\gor(E_n)  -\imat \left(w_\goc+\frac{1}{2}\right) T_\goc(E_n)
\end{equation}
for non-negative integer~$(w_\goc,w_\gor)$ and a fraction of time~$\tau$
 strictly smaller that~$T_\goc(E_n)$ that depends on the initial and final conditions 
(those are not necessarily symmetric). Then we have
\begin{equation}
  \tilde{S}_\goo(b_{\beta'}, b_{\beta^{}},E_n)=
  w_\gor \tilde{S}_\gor(E_n)  +\imat \left(w_\goc+\frac{1}{2}\right) \tilde{S}_\goc(E_n)
\end{equation}
and we take
\begin{equation}
  \mu_\goo=w_\gor+w_\goc+1\;.
\end{equation}
For the same reason as previously explained the dominant contributions  come from
those orbits where~$w_\goc=0$.
In order to mimic a real~$T$, we will choose large winding numbers~$w_\gor$ such 
that~$\re T=w_\gor T_\gor(E_n)\gg\im T=T_\goc(E_n)/2$. Because of the quantization 
condition in the right well~$\tilde{S}_\gor(E_n)=(n+1/2)2\pi\hbar$, the rapid oscillations
$\exp(\imat\re S_\goo/\hbar)$ disappear (or inversely if we want to maintain~$\Delta_n$ real
to first order, we recover the usual quantization condition). Then
we obtain 
\begin{equation}\label{eq:Delta_nGarg}
  \Delta_n\scl \frac{2 \hbar}{T_\gor(E_n)} \, \EXP{-\tilde{S}_\goc(E_n)/(2\hbar)}\;.
\end{equation}
The classical
frequency $1/T_\gor(E_n)$  attached to~$\gor$ is essentially of order~$\omega/(2\pi)$. 
$\Delta_n(T)$ becomes independent of~$T$
for large~$w_\gor$:  the behaviour of~$T$  is
mainly governed by~$w_\gor T_\gor$,  then the $1/T$ prefactor in \eqref{eq:scDeltan}
is compensated by the increasing  number of identical terms in 
the sum since we have seen that the number of  topological classes of orbits
increases linearly with~$4w_\gor$ (the factor~4 comes from the two possible initial branches~$\beta=\pm$
and the two possible final branches~$\beta'=\pm$; in other words from the 
sequences of~$[\rho_1,\rho_2,\dots]$ beginning either by~$q_r$ or $q_r'$ and ending 
either by~$-q_r$ or $-q_r'$). 
The discrepancy between~\eqref{eq:Delta_nGarg} and Garg's formula
is just the factor~$g_n$ given by~\cite[eq. (B2)]{Garg00a} (see also\cite[eq.~(3.41)]{Connor+84a}) that tends to~1 when $n$ increases:
 $g_0=\sqrt{\pi/\mathrm{e}}\simeq1.075$, $g_1\simeq1.028$, \dots
There is also a ratio of order one, more precisely~$\sqrt{2/\pi}\simeq0.8$, between
estimations~\eqref{eq:Delta_0Garg}
and \eqref{eq:Delta_nGarg} taken for~$n=0$; the second is slightly better and  coincides with the 
formula given in Landau and Lifshitz~\cite[sec.~50, problem~3]{Landau/Lifshitz77a}.
Again, these discrepancies
come from the different nature of the approximations that are involved.

Let us end this section by a short comment on the connection with the
usual instanton theory where~$\re T=0$ and where~$\im T\to-\infty$.
This regime, that allows to select the ground-state doublet only, is
included in our approach because the instanton trajectories appear to
be the limit of $\goc$ getting closer to the separatrix~$\goS$ whereas
the classical real oscillations in the wells shrink to the equilibrium
points.  All along this paper we emphasize that the phase-space
representation is particularly appropriate and it is straightforward
to recover the usual picture of instantons (for instance~$q(\imat t)$
versus~$\imat t$) from our figure~\ref{fig:2puits} d).

\section{Dynamical tunnelling for the simple pendulum}\label{sec:pendulum}

\begin{figure}[!ht]
\center
\includegraphics[width=8.5cm]{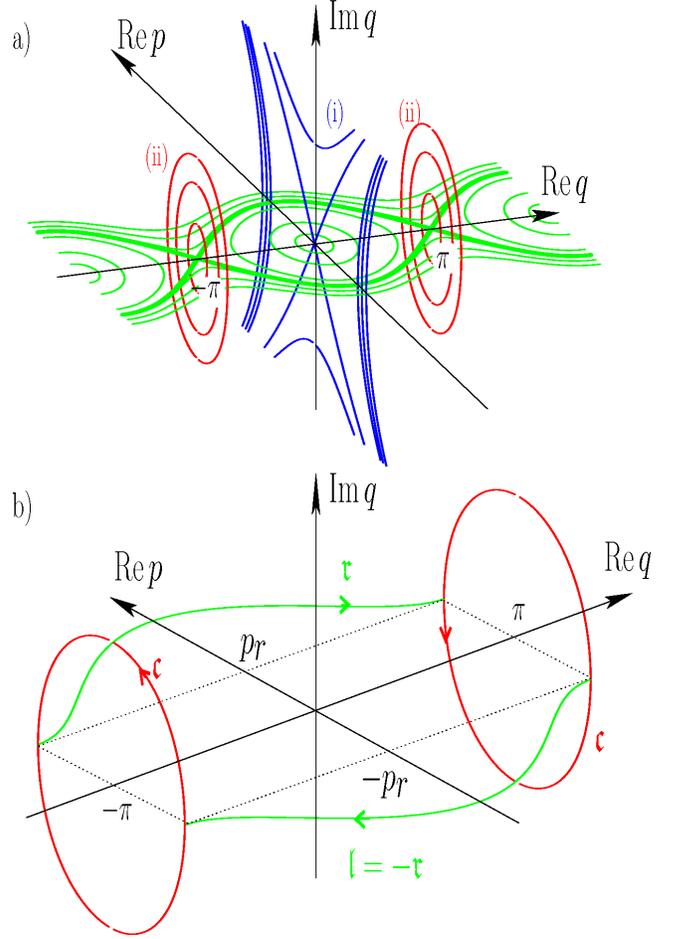}
\caption{\label{fig:pendule} (colour online) 
With the choice of a complex-time path
  given in \ref{fig:slidingstaircase}) (one-step path), the
  contributions to~$\tr\!\big(\opS\,\hat\varPi_n\,\hat{U}(T)\big)$
  come from one half symmetric periodic orbits~$\goo$ (b) that are a
  composite of 1) (repetitions of) a periodic orbit~$\gor$ of
  period~$T_\gor$ in the phase-space plane~$(\re q,\re p)$  at energy~$E>\gamma$ (recall that the planes $q=\pm\pi$ are
  identified); 2)~one half
  of a periodic
  orbit~$\goc$ of period~$T_\goc$ in the phase-space plane~$(\im q,\re
  p)$ at energy~$E$. }
\end{figure}

The simple pendulum corresponds to~$V(q)=-\gamma\cos q$ with~$\gamma>0$
and strictly periodic boundary conditions that identify~$q=-\pi$ and~$q=\pi$.
At energy~$E>\gamma$, the classical rotation with~$p=\sqrt{2(E-V(q))}$
can never switch to the inverse rotation with~$-p=-\sqrt{2(E-V(q))}$. 
At the quantum level, the Schr\"odinger's equation for the stationary wave-function~$\phi$
\begin{equation}\label{eq:schrodingerpendule}
  \left[-\frac{\hbar^2}{2}\frac{\dmat^2 }{\dmat q^2}-\gamma\cos q \right]\phi(q)=E\phi(q)
\end{equation}
 leads to the Mathieu equation~\cite{Abramowitz/Segun65a},
\begin{equation}
  y''(x)+\big(a-2g\cos(2x)\big)y(x)=0
\end{equation}
with $x=q/2$, $y(x)=\phi(2x)$, $a=8E/\hbar^2$ and~$g=-4\gamma/\hbar^2$. The $2\pi$-periodicity of~$\phi$ forces $y$ to be
$\pi$-periodic.
 The eigenfunctions can be classified according
to the parity operator:
Even $\pi$-periodic solutions
exist only for a countably infinite set of characteristic values of~$a$ denoted by $\{a_{2n}\}$
with~$n=0,1,\dots$  
Odd solutions correspond to another set, $\{b_{2n}\}$,
with~$n=1,2\dots$ (the $\{a_{2n+1},b_{2n+1}\}$ correspond to $\pi$-antiperiodic solutions 
 and will be rejected). The discrete energy spectrum corresponding to even and odd
solutions of~\eqref{eq:schrodingerpendule} is then $\{E^+_n=\hbar^2 a_{2n}/8, n=0,1\dots\}$ 
and~$\{E^-_n=\hbar^2 b_{2n}/8, n=1,2\dots\}$
 respectively.
Any eigenstate $\ket{\phi_n^\pm}$ with energy~$E^\pm_n\simeq E_n>\gamma$ has its Husimi 
distribution spread symmetrically between the two half phase-space of 
positive and negative~$p$, near the lines $\pm\sqrt{2(E_n-V(q))}$ that define 
two disconnected tori in the cylindrical phase-space. 
If we prepare a wave-packet localised on the line $p=\sqrt{2(E_n-V(q))}$,
since it is no longer a stationary state, its average momentum
 will oscillate between two opposite values, with a tunnelling 
frequency equal to $\Delta E_n/\hbar$ where
\begin{equation}\label{eq:DeltaE_n_pendule_exact}
\Delta E_n=E^+_n-E^-_n=\frac{\hbar^2}{8}(a_{2n}-b_{2n})
\end{equation}
 is the splitting between
the two quasi-degenerate eigenenergies. 
To compute~$\Delta E_n$, we will use \eqref{eq:DeltaEn} with the operator~$\hat\varPi_n$
very much like the exact projector~$\ket{\phi_n^+}\bra{\phi_n^+}+\ket{\phi_n^-}\bra{\phi_n^-}$: 
Its matrix element $\bra{p'}\hat\varPi_n\ket{p}$ will vanish rapidly as soon as $p$ or $p'$
lie outside the region of the two tori. The main contribution to the semiclassical expansion
of~$\tr\!\big(\opS\,\hat\varPi_n\,\hat{U}(T)\big)$ will come from classical trajectories
that connect two symmetric tori. The trace will be semiclassically computed in the momentum basis
and we will choose the complex time path to maintain~$p$ real.
To construct one trajectory at energy~$E>\gamma$ connecting the two tori requires to have purely  
imaginary $q$ whenever~$-p_r<p<p_r$ with~$p_r\DEFt\sqrt{2(E-\gamma)}$ being the classical turning point 
in momentum. More precisely, $q$ is given by~$\cos q=-1-(p_r^2-p^2)/(2\gamma)$. 
From~\eqref{eq:hamiltoneq_dpds}\,
\begin{equation}
  \frac{\dmat t}{\dmat s}=-\frac{1}{\gamma\sin q}\frac{\dmat p}{\dmat s}\,
\end{equation}
we see immediately that $\sin \big(q(s)\big)$ and $\dmat t/\dmat s$
will be real when~$p(s)>p_r$ or~$p(s)<-p_r$ and purely imaginary
otherwise.  In the latter case, since we keep~$p_2(s)\DEFt\im p(s)=0$,
equations~(\ref{subeq:hamiltoneq2}b,c) lead to two possible families
of solutions (i)~$q_1(s)\DEFt\re q(s)=0$ and (ii)~$q_1(s)=\pm\pi$.
Then, with $p_1\DEFt\re p$ and~$q_2\DEFt-\im q$,
equations~(\ref{subeq:hamiltoneq2}a,d) become
\begin{subequations}
  \begin{align}
  \frac{\dmat p_1}{\dmat s}&=\pm\gamma\sh q_2\left(\imat\frac{\dmat t}{\dmat s}\right)\;;\\
  \frac{\dmat q_2}{\dmat s}&= p_1\left(\imat\frac{\dmat t}{\dmat s}\right)
  \end{align}
\end{subequations}
 and are associated with a real time dynamics governed by the Hamiltonian 
\begin{equation}
  \tilde H(p_1,q_2)=p_1^2/2\mp\gamma\ch q_2\;.
\end{equation}
with ``$-$'' corresponding to case (i) and  ``$+$'' corresponding to case (ii).
Then, instantons correspond to trajectories evolving 
in the transformed potential~$\tilde V(q)\DEFt\mp \gamma\ch q$ rather than the
usual inverted potential~$-V(q)$.
We will choose a one step complex time path as in Fig.~\ref{fig:slidingstaircase}
and we will represent the orbits in a three dimensional space $(\re q, \re p, \im q)$
and the connection between the allowed and the forbidden trajectories occurs
on the axes~$(\re q=0, \im q=0)$  and~$(\re q=\pm\pi, \im q=0)$ (Fig.~\ref{fig:pendule}a).
Trajectories in family (i) escape from the unstable point at $\re p=0,\im q=0$ without coming back to the 
plane $\re p=0,\re q=0$. Only orbits in family (ii) can be used to produce periodic orbits. A typical 
periodic orbit~$\goo$ 
connecting two symmetric rotations of the pendulum is given in Fig.\ref{fig:pendule}b).
We will choose $\-\im T$ to be precisely the half period of the periodic orbit~$\goc$ of family (ii) 
at energy~$E_n$ and  $\re T$ to be (an integer multiple of) the period of rotation 
of the pendulum at the same energy. This choice exhibits the  dominant contribution 
to~$\tr\!\big(\opS\,\hat\varPi_n\,\hat{U}(T)\big)$. We can reproduce the same reasoning that led to
\eqref{eq:Delta_nGarg} with 
the r\^ole of~$p$ and~$q$ being exchanged. 
Now $T_\gor=\dmat \tilde S_\gor/\dmat E$ is the typical frequency
on the torus at energy~$E_n$. 

\begin{figure}[!Ht]
\center
\includegraphics[width=7.5cm]{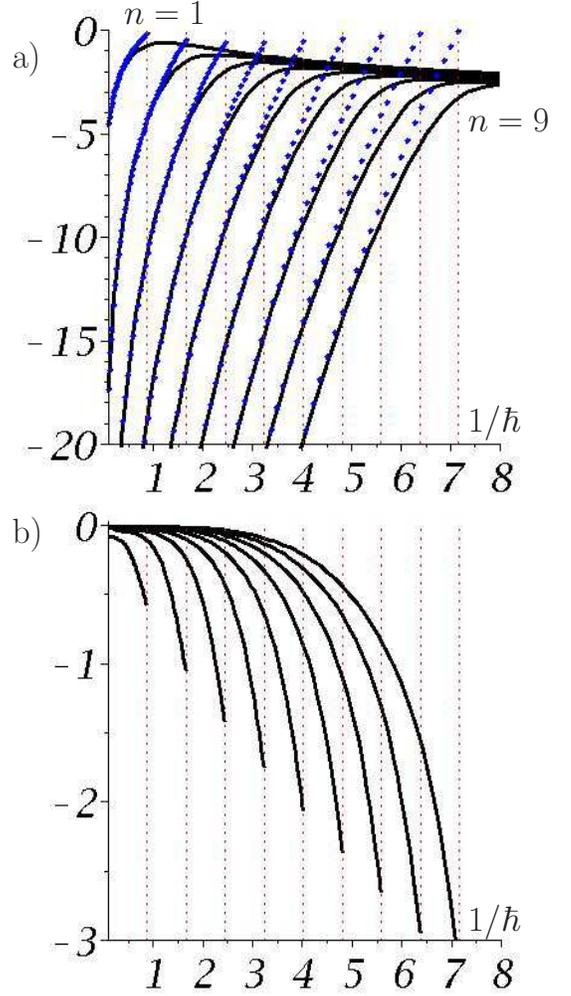}
\caption{\label{fig:Delta_n_pendule} For the pendulum with~$\gamma=1$,
 we have plotted in a)  the exact value of $\ln(\Delta E_n)$ 
(see eq.~\eqref{eq:DeltaE_n_pendule_exact}) versus~$1/\hbar$
 for~$n=1,\dots9$ (black lines). The dots are given by the
semiclassical approximation $\ln(\Delta_n)$ (see eq.~\eqref{eq:Delta_npendule_bis}) up to the maximum value
of $1/\hbar$ (vertical dotted lines) where~$E_n^+$ become lower that the 
separatrix energy~$\gamma$. The semiclassical limit is obtained when
the states become more and more localised in the region of phase space corresponding to rotations: for
a fixed~$n$ this requires to increase~$E_n\simeq n^2\hbar^2/2$ that is to \emph{increase}~$\hbar$; or
for a fixed energy $E>\gamma$, we must increase~$n\propto\sqrt{2E}/\hbar$. 
 To control 
the validity of 
the prefactors, we can check in b) that $\ln(\Delta E_n/\Delta_n)$ approaches zero in the semiclassical limit. }
\end{figure}

Keeping only the contributions that provide a positive imaginary part of the action,
expression \eqref{def:S}
becomes 
\begin{multline}
  S=\int_{s_{i}}^{s_{\!f}}
    p_1(s)\frac{\dmat q_1}{\dmat s}(s)\dmat s -E\re T \\
   +\imat\left(\int_{s_{i}}^{s_{\!f}}
    p_1(s)\frac{\dmat q_2}{\dmat s}(s)\dmat s +E(-\im T)\right).
\end{multline}
Then
\begin{equation}
  S_\goo(E)=\tilde{S}_\goo(E)-E T_\goo
\end{equation}
 with $T_\goo=T$ and
\begin{multline}
  \tilde{S}_\goo(E)=w_\gor\tilde{S}_\gor(E)+\imat\tilde{S}_\goc(E)/2\\
   =2w_\gor\int_0^\pi\sqrt{2(E+\gamma\cos q)}\,\dmat q\\
+2\imat \int_0^{\mathrm{argch}(E/\gamma)}\sqrt{2(E-\gamma\ch q)}\,\dmat q\;.
\end{multline}
All these expressions can be written in terms of the complete elliptic 
integrals~\cite{Gradshteyn/Ryzhik65a}
(defined for~$|u|<1$)
\begin{subequations}
  \begin{align}
\mathcal{K}(u)&\DEF\int_0^{\pi/2}\frac{\dmat x}{\sqrt{1-u^2\sin^2x}}\;;\\
\mathcal{E}(u)&\DEF\int_0^{\pi/2}\sqrt{1-u^2\sin^2x}\;\dmat x\;.
\end{align}
\end{subequations}
Namely,
\begin{subequations} 
\begin{align}
 \tilde{S}_\gor(E)&= 4\sqrt{2(E+\gamma)}\,\mathcal{E}\left(\!\sqrt{\frac{2\gamma}{E+\gamma}}\,\right)\;;\\
  \tilde{S}_\goc(E)&=8\sqrt{2(E+\gamma)}
\left[\mathcal{K}\left(\!\sqrt{\frac{E-\gamma}{E+\gamma}}\,\right)
                      -\mathcal{E}\left(\!\sqrt{\frac{E-\gamma}{E+\gamma}}\,\right)\right].
 \end{align}
\end{subequations}
Then we obtain
  \begin{equation}\label{eq:Delta_npendule}
  \Delta_n \sim \frac{2\hbar}{T_\gor(E_n)}  \, \EXP{-\tilde{S}_\goc(E_n)/(2\hbar)}
\end{equation}
with
\begin{equation}
  T_\gor(E)=\frac{\dmat \tilde S_\gor}{\dmat E}=
   \frac{2\sqrt{2}}{\sqrt{E+\gamma}}\,\mathcal{K}\left(\!\sqrt{\frac{2\gamma}{E+\gamma}}\,\right)\;.
\end{equation}
The energies of the highly excited states are approximately given by the free rotations:
 $E_n\simeq n^2\hbar^2/2\gg\gamma$. Then~$T_\gor(E_n)\simeq 2\pi/(n\hbar)$. The asymptotic expansion
of~$\tilde{S}_\goc(E)$ for large~$E$ leads to
\begin{subequations}\label{subeq:Delta_npendule}
  \begin{eqnarray}
  \Delta_n&\sim&\frac{\hbar^2}{\pi}\frac{\sqrt{2E_n}}{\hbar}\;
    \EXP{2(\ln(\gamma/E_n)+2-3\ln2)\sqrt{2E_n}/\hbar}\;;
 \\\label{eq:Delta_npendule_bis}
    &\sim& \frac{1}{\pi n^{4n-1}}\left(\frac{\mathrm{e}}{2}\right)^{4n} \hbar^2 \left(\frac{\gamma}{\hbar^2}\right)^{2n}\;.
\end{eqnarray}
\end{subequations}
The last expression corresponds exactly to equation~(3.44)
 of~\cite{Connor+84a} obtained with standard uniform semiclassical 
analysis. We see on figure~\ref{fig:Delta_n_pendule}
 that~\eqref{eq:Delta_npendule} is a very good
approximation even when the energies $E_n$ get close to~$\gamma$ the energy of the separatrix.

\section{Resonant tunnelling and Fabry-P\'erot effect}\label{sec:resonanttunnelling}

\begin{figure}[!ht]
\center
\includegraphics[width=8.5cm]{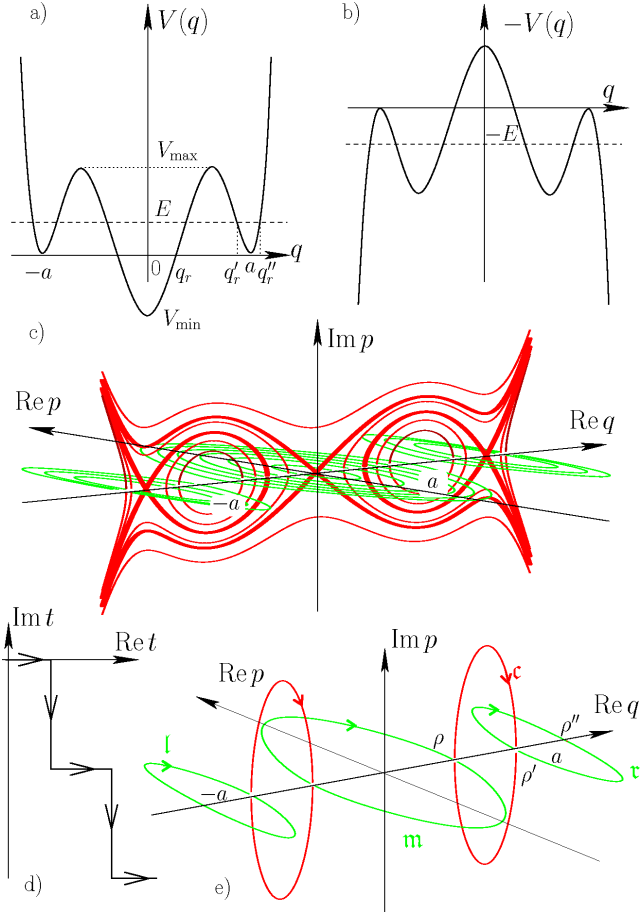}
\caption{\label{fig:3puits} (colour online) To describe resonant tunnelling between two symmetric 
wells centered at~$\pm a$ and separated by a third central well (a), we use 
orbits made of five primitive trajectories with a time path given in (d).
We still maintain the position $q$ to be real and can use a three dimensional section of phase-space
(see text). Here we have chosen $V$ given by~\eqref{eq:Vq6} with $a=7/4$ and $b=1/2$.
} 
\end{figure}

We are now ready the see how formula~\eqref{eq:scDeltan} allows us to reproduce the resonant tunnelling between
two wells related by parity
when the potential in~\eqref{eq:Hp2plusV} has also a deeper central well (Fig.~\ref{fig:3puits}~a).
The minimum of the right and left wells is fixed at zero,  the minimum of the central well 
is~$V(0)=V_\mathrm{min}<0$ and the local maximum between the wells is denoted~$V_\mathrm{max}$.
 When~$0<E<V_\mathrm{max}$, we will denote by~$q_r(E)<q'_r(E)<q_r''(E)$ the three
positive solutions of~$V(q)=E$.
As explained in section~\eqref{sec:doublewell}, we will try to construct appropriate time paths~$[t]$
to exhibit complex trajectories~$\goo$ with purely real~$q$ that connect the two 
symmetric tori from~$\gor$ to~$\gol$ at some energy~$0<E<V_\mathrm{max}$. These tori are delimited by the two
turning points~$\pm\rho'=(0,\pm q_r')$ and~$\pm\rho''=(0,\pm q_r'')$. What is new of course
is the existence of a central real torus~$\gom$ delimited by $\rho=(0,q_r)$ and~$-\rho$.
 Using the three dimensional. 
representation of the section~$\im q=0$ of phase space (Fig.~\ref{fig:3puits}~c and~e), 
we see that the orbits~$\goo$ must be a series of concatenation 
of five trajectories connecting at the turning points~$\pm\rho,\pm\rho'$. 
 First we start with one portion 
living on~$\gor$ with~$\dmat t/\dmat s>0$ and $\im p=0$, then connect at~$\rho'$ to a trajectory
with~$\re p=0$ where~$\imat \dmat t/\dmat s>0$. It follows the 
energy curve~$\goc$ whose equation is~$(\im p)^2/2-V(q)=-E$. 
Then~$\goo$ can connect at~$\rho$ to a real trajectory 
on~$\gom$ with~$\im p=0$ and~$\dmat t/\dmat s>0$, then can cross~$-\goc$ 
from~$-\rho$ to $-\rho'$ before reaching~$\gol=-\gor$. The corresponding time path will necessary 
have at least two steps (Fig.~\ref{fig:3puits}~d) each of them having a height 
which is a half-integer multiple of~$T_\goc$, the real period of the primitive periodic 
orbit~$\goc$. Among all the possible~$\goo$'s, we will keep only 
the exponentially dominant contributions, when~$\goo$ remains as shortly as possible with complex~$p$.
Then, for such orbits to exist, we must choose~$T$ of the form
\begin{equation}\label{eq:T3wells}
  T=\tau+w_\gor T_\gor(E) +\left(w_\gom+\frac{1}{2}\right) T_\gom(E) -\imat T_\goc(E)
\end{equation}
involving the periods of the primitive orbits  and the corresponding
 winding numbers~$w_\gor$, $w_\gom$ which are non-negative integers. $\tau$ denotes 
a positive fraction of time smaller than~$T_\gor(E)$. The base points for 
the two branches~$\beta=\pm$ defined by $\re p\gtrless0$ on~$\gor$ coincide with
the turning point~$b_\pm=q'_r$, the action~$\tilde{S}_\goo(b_{\beta'},b_{\beta^{}},E_n)$  and the index~$\mu_\goo$
are independent of the choice
of the branch where~$\goo$ starts:
\begin{equation}
 \tilde{S}_\goo(b_{\beta'},b_{\beta^{}},E_n)=w_\gor \tilde{S}_\gor(E) +\left(w_\gom+\frac{1}{2}\right) \tilde{S}_\gom(E) +\imat \tilde{S}_\goc(E)\;.
\end{equation}
\begin{equation}
  \mu_\goo=w_\gor +w_\gom+3\;.
\end{equation}
Explicitly, we have
\begin{subequations}\label{def:Stildeq6}
\begin{align}
  \tilde{S}_\gor(E)&\DEF2\int_{q'_r(E)}^{q''_r(E)}\!\!\sqrt{2\big(E-V(q)\big)}\,\dmat q\;;\\
  \tilde{S}_\gom(E)&\DEF4\int_{0}^{q_r(E)}\!\!\sqrt{2\big(E-V(q)\big)}\,\dmat q\;;\\
  \tilde{S}_\goc(E)&\DEF2\int_{q_r(E)}^{q'_r(E)}\!\!\sqrt{2\big(V(q)-E\big)}\,\dmat q\;;
\end{align}
\end{subequations}
and the corresponding periods~$T_\gor$, $T_\gom$, $T_\goc$ are obtained by deriving with respect to $E$.

As we have seen, formula~\eqref{eq:scDeltan} will provide an approximation of the exact 
splitting~$\Delta E_n$  that becomes better, the better the condition  
 $\re T(E_n)\gg\im T(E_n)$ is satisfied. Not only  this condition 
render the precise value of~$\tau$ irrelevant, it also requires large~$w_\gom$ and/or $w_\gor$, especially
if we work with~$E_n\gtrsim 0$ for which~$T_\goc(E_n)\gg T_\gom(E)$  and~$T_\goc(E_n)\gg T_\gor(E)$.
For a given pair of~$w_\gor$ and~$w_\gom$, there are $4(w_\gor+1)$ topological classes of orbits 
corresponding to two possible initial  branches, two possible final branches and~$r=0,\dots, w_\gor$
possible windings on~$\gor$ for~$w_\gor-r$ windings on~$\gol$.  Then all different~$w_\gor$ and~$w_\gom$,
such that~\eqref{eq:T3wells} holds, give a contribution to~\eqref{eq:scDeltan}: If we 
define~$R(T)\DEFt\re T-\tau-T_\gom/2\simeq\re T$ then,
  \begin{multline}\label{eq:scDeltan_3wells}
  \Delta_n(T)\scl\frac{2\hbar}{T}\; \EXP{-\tilde{S}_\goc(E_n)/\hbar}\\
 \times \hspace{-1ex}  \sum_{\substack{\{w_\gor,w_\gom\} \ \text{ pos. int. such that}\\ w_\gor T_\gor(E_n)+w_\gom T_\gom(E_n)=R(T)}}
   \hspace{-4em} (w_\gor+1)\,
\EXP{\imat w_\gor[\tilde{S}_\gor(E_n)/\hbar-\pi]+\imat w_\gom[\tilde{S}_\gom(E_n)/\hbar-\pi]}.
\end{multline} 
\begin{figure}[!t]
\center
\includegraphics[width=8.5cm]{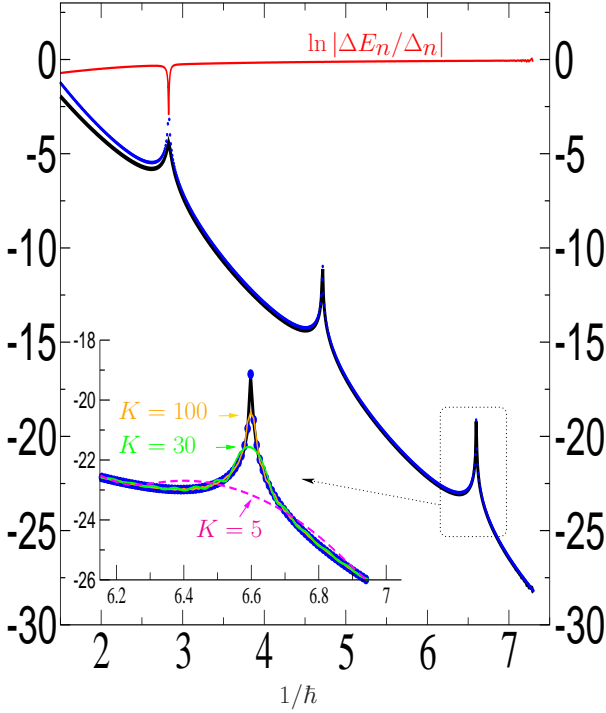}
\caption{\label{fig:comparaisonDeltaEq6}  (colour online) For~$V$ given by~\eqref{eq:Vq6} with $a=7/4$ and $b=1/2$,
we plot $\ln|\Delta E_n|$ (thick black line)  for the lowest doublet in the symmetric wells 
$E^\pm_n\simeq\hbar\omega/2$ with $\omega=2a\sqrt{2(a^2-b^2)}\simeq6.30$. The (blue) dots
correspond to estimation~\eqref{eq:contrib_ppobis} with 
$T=|T|\EXP{-\imat\theta}=(K+1/2)T_\gom(E_n)-\imat T_\goc(E_n)$ in the limit $K\to\infty$ ($\theta\to0$). 
 The upper thin (red) curve
provides $\ln|\Delta E_n/\Delta_n|$. The inset provides a magnification of the third spike 
around~$1/\hbar=6.6$ for which $T_\gom(E_n)=2T_\gor(E_n)\simeq1.58$ and $T_\goc(E_n)\simeq1.50$.
Here, we have also plotted $\ln|\Delta_n(T)|$ for $K=100$ (thin orange (light gray), $\theta\simeq0.01$) $K=30$ 
(thin green $\theta\simeq0.03$), $K=5$ (dashed thin magenta $\theta\simeq0.17$).
}
\end{figure}
We immediately see the resonance at work since the sum reaches a maximum when 
both $\nu_\gor\DEFt\tilde{S}_\gor(E_n)/(2\pi\hbar)-1/2$ 
and~$\nu_\gom\DEFt\tilde{S}_\gom(E_n)/(2\pi\hbar)-1/2$ are integers: the energy of a state 
mainly localised in the central well 
becomes nearly degenerate (up to $\hbar^2$ terms) with the energy doublet in the lateral wells. Then
the contributions of the repetitions of~$\gom$ interfere constructively like the optical rays 
in a Fabry-P\'erot interferometer\cite[secs.~12.14--12.17]{Bohm51a}.
To estimate the sum in the right hand side of~\eqref{eq:scDeltan_3wells}, let us take
a rational approximation of the ratio~$T_\gom/T_\gor$, namely
\begin{equation}\label{eq:ratioofT}
  \frac{T_\gom}{T_\gor}\simeq\frac{m}{r}
\end{equation}
with~$m$ and~$r$ being coprimes positive integers.
 For the polynomial potential
\begin{equation}\label{eq:Vq6}
  V(q)=(q^2-a^2)^2(q^2-b^2)
\end{equation}
with~$a>b>0$, the argument presented in \cite{Khuatduy/Leboeuf93a} can
be generalised to show that~\eqref{eq:ratioofT} is actually exact
with~$r=1$ and~$m=2$ for any energy~$0<E<V_{\mathrm{max}}$. If $K$
denotes the integer part of~$R(T)/(rT_\gom)$, we can compute and
approximate for~$K\gg1$ the right hand side
of~\eqref{eq:scDeltan_3wells} and obtain
\begin{equation}\label{eq:scDeltan_3wells_Kinfinite}
  |\Delta_n(T)|\scl\frac{\hbar}{T_\gor}\frac{\EXP{-\tilde{S}_\goc(E_n)/\hbar}}{\big|\sin\big(\pi(m\nu_\gor-r\nu_\gom)\big)\big|}\;.
\end{equation}

Fig.~\ref{fig:comparaisonDeltaEq6} 
 shows that this latter expression provides a good approximation for~$\Delta E_n$ even
in the immediate neighbourhood of a resonance where estimation~\eqref{eq:DeltaEn} is not justified any more.
If we had continued working with a finite~$K$, the sum \eqref{eq:scDeltan_3wells} would 
have involved a finite number of terms and
the singularities due to the vanishing denominators in~\eqref{eq:scDeltan_3wells_Kinfinite} would have
 been smoothed down (inset in Fig.~\ref{fig:comparaisonDeltaEq6}). In other words, 
for a fixed~$|T|$, rotating down~$T$ 
in the lower half plane~$\im T<0$, destroys very quickly ($\theta\simeq0.2$)
 the  large resonant fluctuations of tunnelling. 
This effect has already been 
shown in the case of a kicked system \cite{Mouchet07a}.

\section{Escape rates}\label{sec:escape_rate}
\begin{figure}[!ht]
\center
\includegraphics[width=8.5cm]{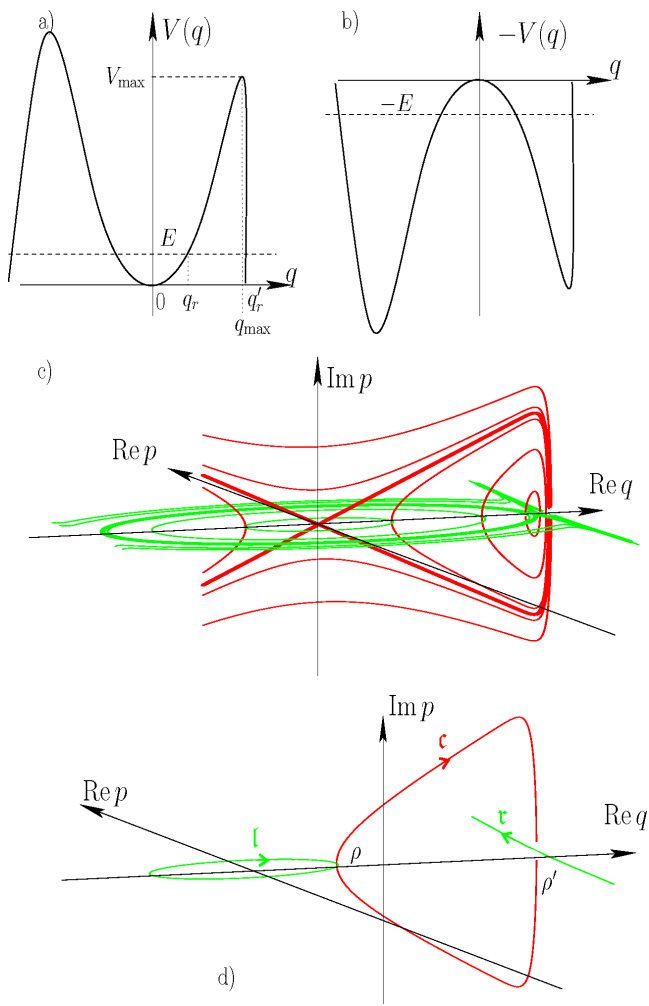}
\caption{\label{fig:sortieilot} (colour online) The escape rate at energy~$E$
 from an island created by a confining potential $V$
whose shape is shown in a) can be semiclassically computed in terms of complex classical orbits
where one canonical coordinate, say~$q$, is kept real.  In phase-space, this trajectory
  appears to be a concatenation of two types of curves that join on
  the $(\re q)$-axis: 1)~a trajectory that lies in the phase-space
  plane~$(\re q,\re p)$ with real variations of $t$ at energy~$E$ with the
  potential~$V$ and 2)~a trajectory that lies in the phase-space
  plane~$(\re q,\im p)$ with imaginary variations of $t$ at energy~$-E$ with the
  potential~$-V$ shown in b).  In c) a family of constant energy curves is
  shown (horizontal green for the first type, vertical red for the
  second type).  In d) for a given energy, we show how three
  curves~$\gol,\goc,\gor$ glue together at the turning points.
    }
\end{figure} 

So far we have focused our analysis of tunnelling in bounded systems
only, but the philosophy we presented here can be extended to more
general situations. For instance, let us show how we can compute the
escape rate from a metastable state localised in a confining
potential~$V$ whose shape has the form given in
Fig.~\ref{fig:sortieilot}~a). The potential has a local minimum
at~$q=0$ (say $V_{\textrm{min}}=0$) and an energy barrier
for~$0<q<q_{\textrm{max}}$ whose height is~$V_\mathrm{max}$.
For~$q>q_{\textrm{max}}$, the potential remains non-positive and
therefore in real phase-space ($\re p$, $\re q$), $V$ defines around
the origin an island of stability made of tori with positive energy.
One state whose Husimi distribution is initially localised in the
island, say a quasi-mode $\ket{\Phi_n}$ at energy $0 <E_n
<V_\mathrm{max}$, will progressively decay outside the well. The decay
rate~$\Gamma_n$ is then defined from the overlap:
\begin{equation}
 \bra{\Phi_n}\hat{U}(T) \ket{\Phi_n}=\EXP{-\Gamma_n T/2-\imat E_n T/\hbar}\;.
\end{equation}
If we choose a complex~$T$ such that
\begin{equation}
  |\Gamma_n T|\ll1\;,
\end{equation}
then we obtain a trace formula for~$\Gamma_n$ with the help of the projector-like operator~$\hat\varPi_n$
\begin{equation}\label{eq:Gamman}
  \Gamma_n\simeq-\frac{2}{\im T}\im\left(\EXP{\imat E_nT/\hbar }\tr \big(\hat\varPi_n\hat{U}(T)\big)\right)
\end{equation}
which allows an explicit semiclassical expansion in terms of classical
solutions with a complex time path.  The dominant contributions will
be provided by periodic orbits in time~$T$ with real~$q$ starting on
the torus at energy~$E_n$, then, while $\imat \dmat t/\dmat s>0$, going
forth outside the well before coming back to its initial starting
point (Fig.~\ref{fig:sortieilot}~c, d)).  Then, up to a dimensionless
factor~$f_n$ of order one, we get
\begin{equation}
  \Gamma_n\simeq \frac{f_n}{T_\goc(E_n)}\EXP{-\tilde{S}_\goc(E_n)/\hbar}
\end{equation}
with
\begin{equation}\label{def:Stildeilot}
  \tilde{S}_\goc(E)\DEF 2\int_{q_r(E)}^{q'_r(E)}\!\!\sqrt{2\big(V(q)-E\big)}\,\dmat q
\end{equation}
where~$0<q_r(E)<q_r'(E)$ are the two right turning points at
energy~$E$. Here we have supposed that the action between the two left
turning points is larger than~\eqref{def:Stildeilot} and gives a
subdominant contribution. If the potential is symmetric, two symmetric
complex orbits would contribute with the same weight and therefore the
escape rate should twice as large.  In the case of an island with
sharp boundaries around~$q_{\textrm{max}}$, $\tilde{S}_\goc(E)$ which
is the area enclosed by the primitive orbit~$\goc$
(Fig.~\ref{fig:sortieilot}~d) is mainly given by the portion of~$\goc$
where we can keep an harmonic approximation for the
potential:~$V(q)\sim\omega^2q^2/2$, $q_r(E)\simeq\sqrt{2E}/\omega$,
$q'_r(E)\simeq\sqrt{A/(\pi\omega)}$ where $A$ is the area of the
island in the real phase space.  Then, with~$a\DEFt \omega A/(2\pi
E)$, we have
\begin{equation}
  \tilde{S}_\goc(E)\simeq
\frac{2E}{\omega}\left[\sqrt{a}\sqrt{a-1}-\ln\big(\sqrt{a}+\sqrt{a-1}\,\big)\right]\;.
\end{equation}
Inserting this expression in~\eqref{eq:Gamman} with
$E_n\simeq(n+1/2)\hbar\omega$, we exactly recover the expression~(5)
used in~\cite{Backer+08a} with an elegant and simple
interpretation.  The chaotic sea that surrounds the integrable island
in the mixed system considered by B\"acker \textit{et al.} acts as a
sharp effective potential barrier as the one draw in
picture~\ref{fig:sortieilot}; our complex trajectory that allows to
escape from the regular region has its main features governed by the
integrable (and even harmonic) approximation of the dynamics about the
island, following precisely the general philosophy
of~\cite{Backer+08a} and~\cite{Lock+09a}. This computation of the ``direct''
tunnelling (by opposition to resonant tunnelling where the model 
of a pure quadratic kinetic energy fails) can also be reproduced
within the standard one-dimensional JWKB theory used for computing 
transmission coefficients.

Here again, we can check easily that the traditionnal instanton method
is included in our approach : The regime $\Re T\to0$ selects
the instanton solution and provides the escape rate from the 
equilibrium point as given by equation (2.47) of \cite[chap.~7]{Coleman85a}.

\section{Conclusions}

The explicit semiclassical expansions of trace formulae for tunnelling
splittings (or escape rates) in terms of classical orbits constructed
with complex-time paths provide an interesting alternative approach to
Herring formulae essentially because they do not require to
analytically continue the wave functions in the complex plane. In
multidimensional tunnelling and/or for a non-autonomous Hamiltonian
system, the generic lack of constants of motion isolates the stable
islands (if any) from each other by chaotic seas.  The analytic
continuation of the \textsc{kam} tori that build the islands is
prevented by the existence of natural boundaries (see for instance the
recent discussion in \cite[part II, appendix~B]{Shudo+09a} and
references therein).  The approach we have presented here seems to
circumvent these difficulties but, of course, the problem of
how to select, with an appropriate $[t]$, the relevant trajectories among an
 a priori exponentially growing number of classical complex solutions
 remains open.
 We expect the tunnelling splittings
between two states at energy around~$E$ localised in two symmetric
islands to be approximated by the expansion of the form
\begin{multline}\label{eq:scDeltan_Nwells}
  \big|\frac{f\hbar}{T}
   \sum_{[\rho_1,\rho_2,\dots]} 
\EXP{\imat \sum_{i=1}^N w_i[\tilde{S}_i(E_n)/\hbar-\pi]}
\EXP{-\sum_{j=1}^M w'_j[\tilde{S}'_j(E_n)/\hbar]}
 \big|
\end{multline}
where the sum runs over all possible sequences of turning
points~$[\rho_1,\rho_2,\dots]$ at energy~$E$ such that we can choose a
complex-time path that leads to a trajectory, made of primitive
orbits, that connects in time~$T$ the two (real) tori.  The
winding~$\{w_i\}$ and the actions~$\{S_i\}$ (resp.~$\{w'_j\}$ and
$\{\tilde{S}'_j\}$) refer to the primitive orbits obtained when the
variations~$\dmat t/\dmat s$ are purely real (resp. purely imaginary).
For dimensions larger that one, the dimensionless prefactor~$f$ may
appear as a power law in $\hbar$.  Here, inspired by the study in
section~\ref{sec:resonanttunnelling}, we can qualitatively see how the
constructive interferences between repeated paths emerge in a
speckle-like forest because of the presence of resonances. As shown in
\cite{Mouchet07a}, a progressive complex rotation of time provides a
natural way to select the main resonance effects. If we want to expand
the splittings (or the escape rates) according to elementary process
as proposed in \cite{Lock+09a}, our approach offers a promising tool
to interpret and compute semiclassically all of the ingredients of
such an expansion.

\section*{Acknowledgments}
We have received a lot of benefits from discussions with Olivier Brodier, Dominique Delande, Akira Shudo, Denis Ullmo and Jean 
Zinn-Justin, with special
thanks to Stephen Creagh, who has had a good intuition about these issues for a
long time and  has shared his insights with us. We acknowledge
Stam Nicolis for his careful reading of the manuscript and Olivier Thibault for his efficient skills in terms of computer
maintenance.
 One of us (A.M.) 
is grateful to the Laboratoire Kastler Brossel for its hospitality.

\appendix
\section{}\label{app:contrib_sc}

In this appendix, we derive the dominant contributions~\eqref{subeq:contrib_gene} 
to the traces involved in~\eqref{eq:DeltaE0}.
Both cases $\tr\!\big(\hat{U}(T)\big)$ and $\tr\!\big(\opS\,\hat{U}(T)\big)$ will be treated simultaneously
by defining the sign~$\eta$ to be $+1$ in the first case and~$-1$ in the second case. The semiclassical arguments
underpinning the derivation are relatively standard and may be found in one way or another in
the literature. For instance, the contribution~\eqref{eq:contrib_ppobis}
for~$\eta=1$ can be found in~\cite[eq.~(2.12)]{Dashen+74a} within a more restricted context
(see also \cite[chap.~8]{Haake01a}). Nevertheless we found it 
useful to provide all the steps in the precise context
of this work, not only to render the presentation self-contained, but also because we are 
working in the time domain with general Hamiltonians that have not necessarily the form~\eqref{eq:Hp2plusV}.

Given a~$[t]$, for any classical phase-space path~$\goo$, we can consider the final coordinates $(p_f,q_f)$ at time~$t(s_f)=T$ 
as smooth functions of the initial coordinates $(p_i,q_i)$ at time~$t(s_i)=0$.  
The monodromy matrix~$M_\goo$ is defined as the differential of these functions: to first 
order, an initial small perturbation implies the final perturbation
\begin{equation}
  \binom{\delta p_f}{\delta q_f}=\begin{pmatrix}M_{\goo,11} & M_{\goo,12}\\
                                                M_{\goo,21} & M_{\goo,22}
                                  \end{pmatrix}
  \binom{\delta p_i}{\delta q_i}\;.
\end{equation}
The sub-matrices can be expressed from the second derivatives of~$S_\goo(q_f,q_i;T)$ by differentiating equations~\eqref{subeq:pipfS}:
\begin{subequations}\label{subeq:MS}
  \begin{align}
    M_{\goo,11} &= -\partial^2_{q_fq_f}S_\goo\left(\partial^2_{q_fq_i}S_\goo\right)^{-1} ;\\
    M_{\goo,12} &=  \partial^2_{q_fq_i}S_\goo
                   -\partial^2_{q_fq_f}S_\goo\left(\partial^2_{q_fq_i}S_\goo\right)^{-1}\partial^2_{q_iq_i}S_\goo\; ;\\
    M_{\goo,21} &= -\left(\partial^2_{q_fq_i}S_\goo\right)^{-1};\\
    M_{\goo,22} &= -\left(\partial^2_{q_fq_i}S_\goo\right)^{-1}\partial^2_{q_iq_i}S_\goo\;;
  \end{align}
\end{subequations}
provided that~$\partial^2_{q_fq_i}S_\goo$ is invertible. Equations~(\ref{subeq:MS}a,c,d) can be inverted in
\begin{subequations}\label{subeq:SM}
  \begin{align}
    \partial^2_{q_fq_i}S_\goo &=-(M_{\goo,21})^{-1}\; ;\\
    \partial^2_{q_iq_i}S_\goo &=(M_{\goo,21})^{-1}M_{\goo,22}\;; \\ 
    \partial^2_{q_fq_f}S_\goo &=M_{\goo,11}(M_{\goo,21})^{-1}\;;  
  \end{align}
\end{subequations}
while~(\ref{subeq:MS}b) provides
\begin{equation}
  M_{\goo,12}=-(M_{\goo,21})^{-1}+M_{\goo,11}(M_{\goo,21})^{-1}M_{\goo,22}
\end{equation}
which is nothing but the expression that~$\det M_\goo=1$ once we use the identity
\begin{equation}\label{eq:detABCD}
  \det\begin{pmatrix}A & B\\ C & D \end{pmatrix}=\det(CAC^{-1}D-CB)
\end{equation}
where~$(A,B,C,D)$ are square matrices of the same size and~$C$ is invertible.

Coming back to the oscillating integral~\eqref{eq:intq}, if the integration path can be deformed in order to pass through an
isolated critical point $q_c$ of~$q\mapsto S_\goo(\eta q,q,T)$, we
will have a contribution of the form
\begin{equation}\label{eq:contrib_pisolated}
\frac{\displaystyle\sqrt{\det\left(\frac{\partial^2 S_\goo}{\partial q_i\partial q_f}
                             \right)
                        }
     }
     {\displaystyle\sqrt{\det\left(-\frac{\partial^2\big(S_\goo(\eta q, q;T)\big)}{\partial q\partial q} 
                             \right)
                        }
     }\;
  \EXP{\imat S_\goo/\hbar}
\end{equation}
where all the functions are evaluated at~$(\eta q_c,q_c,T)$. Physically $q_c$ is interpreted as the initial position
of the phase space paths~$\goo$ such that $(p_f,q_f)=(\eta p_i,\eta q_i)$.  With
the help of~\eqref{subeq:SM}, the prefactor of the exponential can be
written
as~$\big[\det\big(M_{\goo,22}+M_{\goo,21}M_{\goo,11}(M_{\goo,21})^{-1}-2\eta\big)\big]^{-1/2}$
up to a  global sign.  Using~\eqref{eq:detABCD} again,
the contribution~\eqref{eq:contrib_pisolated} can be written as
\begin{equation}\label{eq:contrib_pisolatedbis}
  \frac{(-1)^{\nu_\goo}}{\sqrt{(-\eta)^\textsc{d}\det(1-\eta M_\goo)}}\;\EXP{\imat S_\goo/\hbar}
\end{equation}
where $(-1)^{\nu_\goo}$ fixes the sign of
the square root  and~$\textsc{d}$ demotes the number of degrees of freedom.
Recall that in the definition of the
path-integrals~\eqref{eq:pathint}, a time-slice~$\tau$ is always
implicit.  Let us keep for a moment an explicit discretization, for
instance with $T$ being an integer multiple of~$\tau$, $\goo$ referring
to a discrete set of  points, Hamilton's
equations~\eqref{subeq:hamiltoneq1} being discretized into a
phase-space map, path integral~\eqref{def:S} being turned into a discrete (Riemann)
sum, etc. Then, the contribution of each of the points of a given~$\goo$, 
such that~$(p_f,q_f)=(\eta p_i,\eta q_i)$, is the same and remains
given by \eqref{eq:contrib_pisolatedbis}; but this is correct only if,
for a given~$\tau$, $\hbar$ is small enough,  because when performing the
steepest-descent method on \eqref{eq:intq}, one must be able to 
split the oscillating integral into separate contributions
coming from two distinct points of~$\goo$. 
In the continuous time limit, \textit{i.e.} when the limit~$\tau\to0$ is taken before
the semiclassical limit~$\hbar\to0$ (see [??]), the orbits~$\goo$ of non-zero length
appear as a one-dimensional continuum  none of whose points can be considered separately anymore.
The only isolated critical points~$q_c$ are
given by the position of some equilibrium points. Under the symmetric 
condition~$(q_i,p_i)=(-p_i,-q_i)$, only the origin must be examined.
Linearising the Hamiltonian flow about a non-degenerated fixed point~$\goe=(p_e,q_e)$
leads to a monodromy matrix whose eigenvalues can be collected by pairs 
$\exp(\pm\lambda_{\goe,\alpha} T)$ where~$\{\lambda_{\goe,\alpha}\}_{\alpha=1,\dots\textsc{d}}$
 are the Lyapunov exponents. Then 
\eqref{eq:contrib_pisolatedbis} becomes
\begin{equation}\label{eq:contrib_ebis}
  \frac{(-1)^{\nu_\goe}\,\EXP{-\imat H(p_e,q_e)T/\hbar}}
       {\prod_{\alpha=1}^{\textsc{d}}
           (\EXP{\lambda_{\goe,\alpha} T/2}-\eta\EXP{-\lambda_{\goe,\alpha} T/2})}
\end{equation}
with a possible adjustment of the sign.
 For a generic choice of~$T$, the denominator does not vanish. 

For a non-zero length path~$\goo$,
 the critical~$q$'s are degenerate along the trajectory;
for a system with several degrees of freedom, one must treat separately the (gaussian) integrals
on the transverse  coordinates~$q_\perp$ along which~$S_\goo(\eta q, q, T)$ varies (quadratically)
from the longitudinal coordinates~$q_\parallel$ along which~$S_\goo(\eta q, q, T)$ is constant.
Of course, the dimensions of~$q_\perp$ and $q_\parallel$ depend crucially on the presence of \textsc{kam} tori.
However, multidimensional tunnelling is beyond the scope of this paper and, the
quantitative studies presented
here concern one-dimensional
systems only. The contribution to the trace of such a path is then, up to a global sign,
\begin{equation}\label{eq:intqbis}
   \frac{\EXP{\imat S_\goo/\hbar}\;}{\sqrt{-2\imat\pi\hbar}}
   \int\dmat q \sqrt{\frac{\partial^2 S_\goo}{\partial q_i\partial q_f}
                \bigg|}_{(\eta q, q; T)}.
\end{equation}
The conservation of energy along~$\goo$, $H(p,q)=E$, implicitly defines a function~$p(q,E)$
in the neighbourhood of any point where~$\partial_pH\neq0$. Globally along the trajectory~$\goo$,
we may encounter several possible branches~$p_\beta(q,E)$ for the graph of these functions
(the two possible signs of a square root when the Hamiltonian has the form~\eqref{eq:Hp2plusV}) 
which become singular 
but pairwise connect smoothly at the turning points, defined by $\partial_pH=0$.
  A relation between $E$, $T$, $q_f$ and~$q_i$
can be obtained by integrating~$\int_{s_i}^{s_f}(\dmat t/\dmat s)\, \dmat s$ and 
using \eqref{eq:hamiltoneq_dqds}:
\begin{equation}\label{eq:TEqiqf}
  T=T_\goo(q_f,q_i;E)\DEF\int_{s_i}^{s_f}\frac{\dmat q/\dmat s}{\partial_p H\Big(p\big(q(s),E\big),q(s)\Big)}\;\dmat s\;.
\end{equation}
Everywhere but at the turning points, the value of~$s$ dictates the choice of the branch used for the integrand.
This relation
 implicitly defines~$E_\goo(q_f,q_i;T)$.  
The usual expression for the derivative of implicit functions leads to the relations: 
$\partial_{\!E}\, p\mathop{=}1/\partial_pH$ 
and $\partial_{q_i}E_\goo\mathop{=}-\partial_{q_i}T_\goo/\partial_{E} T_\goo\mathop{=}\partial_T E_\goo/\partial_pH(p_i,q_i)$.
 If we differentiate~\eqref{eq:pf} 
with respect to $q_i$ when $p_f$ is given by~$p\big(q_f,E_\goo(q_f,q_i,T)\big)$, 
 we obtain
\begin{equation}\label{eq:SdEdT}
  \frac{\partial^2 S_\goo}{\partial q_i\partial q_f}\bigg|_{(q_f, q_i; T)}
  =\frac{1}{\partial_p H(p_f,q_f)}\frac{1}{\partial_p H(p_i,q_i)}\frac{\partial E_\goo}{\partial T}\bigg|_{(q_f, q_i; T)}\;.
\end{equation}
With property~\eqref{eq:Hparity}, we get $\partial_pH(\eta p,\eta q)=\eta\partial_pH(p,q)$
and then, by differentiating~\eqref{eq:TEqiqf}, we have $\partial_q\big(T_\goo(\eta q, q, E)\big)=0$. Therefore, the 
energy~$E_\goo$ 
of the path~$\goo$ depends only on~$T$, not on its starting point~$q$. The square root of
$\partial_T E_\goo(\eta q, q, T)=\dmat E_\goo/\dmat T$ can be got out from the integral in~\eqref{eq:intqbis}.
When adding the contribution of each path whose starting point lie on the branch~$\beta$, we obtain
\begin{equation}\label{def:primitiveperiod}
  T_\beta(E)\DEF\int \frac{\dmat q}{\partial_p H\Big(p_\beta\big(q,E\big),q\Big)}\;.
\end{equation}
This is not exactly the right hand side of~\eqref{eq:TEqiqf} because the domain of integration 
in~\eqref{def:primitiveperiod} is the domain of the branch where the starting point of $\goo$ lives.
Each branch~$\beta$ is delimited by two turning points and $T_\beta$
 is the time spent to go from one point to the other.

The integral~\eqref{eq:intqbis} involves all the possible starting points for a trajectory~$\goo$
and therefore we must add all the branches
that patchwork smoothly in phase-space to form the geometrical set of points  crossed by~$\goo$.
 Referring to the purely geometrical quantities 
(\textit{i.e.} independent of the choice
 of the parametrization), we have the contribution 
\begin{equation}\label{eq:contrib_ppobis}(-1)^{\mu_\goo}
  \frac{\left(\sum_\beta T_\beta\right)}{\sqrt{-2\eta\imat\pi\hbar}}
  \sqrt{\frac{\dmat E_\goo}{\dmat T}}\;\EXP{\imat S_\goo/\hbar}
\end{equation}
only if $T_\goo=T$; for  $\eta=+1$ the path $\goo$ is a periodic orbit
and, for $\eta=-1$, the path $\goo$ is half a symmetric periodic 
orbit (the whole periodic orbit being 
of period equal to~$2T$). The sum concerns all the geometrical branches~$\beta$
crossed by~$\goo$ (even if $\goo$ passes several times by the same points, 
each branch is only counted once).
As before, the conversion
of a  product of  two square roots of complex numbers to the
square root of the product  may
introduce a sign that can be absorbed in the definition of~$\mu_\goo$; 
 the exact computation of the index~$\mu_\goo$ is difficult but since it may change
at the bifurcation points only, where the semiclassical approximation fails, it is 
sufficient to know that it depends on the nature and the number
of the turning points encountered on~$\goo$. Therefore it is an additive quantity when
several primitive orbits are repeated or concatenated together.

\section{}\label{app:Sasymptotic}

In this appendix we explain how to obtain the 
asymptotic expansions~\eqref{subeq:Sasymptotique} as $E\to0^+$.

First consider $\tilde{S}_\gor$ and split it in two parts $\tilde{S}_++\tilde{S}_-$ where
\begin{subequations}
\begin{eqnarray}
 \tilde{S}_+&\DEF&\int_a^{q_r'(E)}\!\!2\sqrt{2\big(E-V(q)\big)}\,\dmat q\;;\\
\tilde{S}_-&\DEF&\int_{q_r(E)}^a\!\!2\sqrt{2\big(E-V(q)\big)}\,\dmat q\;.
\end{eqnarray}
\end{subequations}
Setting~$\epsilon\DEF a-q_r(E)$, rewrite~$\tilde{S}_-$ as
\begin{equation}
  \tilde{S}_-=2\sqrt{2}\epsilon\int_0^1\sqrt{V(a-\epsilon)-V(a-s\epsilon)}\,\dmat s,
\end{equation}
Now expand the integrand as a power series in $\epsilon$ up to the fourth order, compute the integrals that appear in each coefficient
 and insert the expansion of $\epsilon$ in~$E$ obtained from the implicit equation~$V(a-\epsilon)=E$:
\begin{multline}\label{eq:epsilon}
  \epsilon=\frac{\sqrt{2}}{\omega}\,\sqrt{E}+\frac{V^{(3)}(a)}{3\omega^4}\,E\\
-\frac{\sqrt{2}\Big(3\omega^2V^{(4)}(a)-5\big(V^{(3)}(a)\big)^2\Big)}{36\omega^7}\,E^{3/2}+\mathrm{O}(E^2)\;.
\end{multline}
Proceed in an analogous way for the computation of the first three terms of the asymptotic expansion in~$\sqrt{E}$ for $\tilde{S}_+$.
When summing~$S_+$ and~$S_-$, expression~\eqref{eq:Srasymptotique} is obtained with~\eqref{def:B}. 

The expansion of $\tilde{S}_\goc$ is more subtle since it is not differentiable at~$E=0$. Its derivative is given by
\begin{equation}\label{eq:dSdE}
  \frac{\dmat \tilde{S}_\goc}{\dmat E}=-\int_\epsilon^a L(s,\epsilon)\,\dmat s
\end{equation}
where we denote~$\epsilon\DEF a-q_r(E)$ and define 
\begin{equation}
  L(s,\epsilon)\DEF\frac{2\sqrt{2}}{\sqrt{V(a-s)-V(a-\epsilon})}\;.
\end{equation}
The function~$L(s,\epsilon)$ is not continuous but we can extract the discontinuous part from 
\begin{equation}
  L(s,\epsilon)=\frac{4}{\omega\sqrt{s^2-\epsilon^2}}\left(\!\!1+\!\!\sum_{n\geqslant3}\frac{(-1)^n2V^{(n)}(a)}{n!}\frac{s^n-\epsilon^n}{s^2-\epsilon^2}\!\right)^{\!\!\!\!-\frac{1}{2}}
\end{equation}
by expanding the last factor:
\begin{equation}\label{eq:L}
  L(s,\epsilon)=\frac{4}{\omega\sqrt{s^2-\epsilon^2}}+\frac{2V^{(3)}(a)}{3\omega^3}\frac{s^3-\epsilon^3}{(s^2-\epsilon^2)^{3/2}}
+M(s,\epsilon)
\end{equation}
where now~$M(s,\epsilon)$ is a continuous function of its two variables. Then, a standard theorem in analysis assures 
that~$\epsilon\mapsto\int_\epsilon^aM(s,\epsilon)\,\dmat s$ is continuous and its limit when~$\epsilon\to0$ is
\begin{eqnarray}
  \int_0^aM(s,0)\,\dmat s&=&\int_0^a\left( L(s,0)-\frac{4}{\omega s}-\frac{2V^{(3)}(a)}{3\omega^3}\right)\dmat s\notag\\
  &=&\frac{4A}{\omega} -\frac{2aV^{(3)}(a)}{3\omega^3}
\end{eqnarray}
with $A$ given by \eqref{def:A}. The two other 
integrals obtained by inserting \eqref{eq:L} in the right hand side of 
\eqref{eq:dSdE} can be computed exactly and  
expanded as~$\epsilon\to0^+$ up to order~$\mathrm{o}(1)$. Then, 
inserting~\eqref{eq:epsilon}, we obtain
\begin{equation}
  \frac{\dmat \tilde{S}_\goc}{\dmat E}=\frac{4}{\omega}\ln\left(\frac{\sqrt{2E}}{2a\omega}\right)-\frac{4A}{\omega}+\mathrm{o}(1)\;.
\end{equation}
Its integration leads directly to \eqref{eq:Scasymptotique}.

\section{}\label{app:contrib_sc_Pin}

The quasi-mode~$\ket{\Phi_n}\DEFt(\ket{\phi_n^+}+\ket{\phi_n^-})/\sqrt{2}$ is localised on one torus at 
energy~$E^+_n\simeq E^-_n$. Standard \textsc{jwkb} techniques \cite{Keller58a,Percival77a} provide 
a semiclassical
approximation to its wave function:
\begin{multline}\label{eq:quasimodesc}
  \Phi^{\mathrm{s.c.}}_n(q)=\frac{1}{\sqrt{\sum_\beta T_\beta(E_n)}}
\sum_\beta  \frac{A_{n,\beta}}{\sqrt{\partial_p H\big(p_\beta(q,E_n),q\big)}}
\\ \times \exp{\left( \imat \int^q_{b_\beta} p_\beta(x,E_n)\,\dmat x/\hbar\right)}
\end{multline}
 ($\beta$ labels the possible several branches of the torus, $A_{n,\beta}$ are
dimensionless coefficients of unit modulus, ${b_\beta}$ is a base point of the 
branch~$\beta$ and~$T_{\beta}(E_n)$ the characteristic time~\eqref{def:primitiveperiod}
spent on the branch $\beta$).
Within the semiclassical approximation, it  is, therefore, consistent
to construct $\varPi_n(q',q)$ by substituting \eqref{eq:quasimodesc} in the matrix elements of
 the projector operator
$\ket{\phi_n^+}\bra{\phi_n^+}+\ket{\phi_n^-}\bra{\phi_n^-}
=\ket{\Phi_n}\bra{\Phi_n}+\opS\ket{\Phi_n}\bra{\Phi_n}\opS$.
From the integral~\eqref{eq:intqqprimePin} 
\begin{equation}
  2\int\dmat q\,\dmat q'\; \Phi_n^{\mathrm{s.c.}}(q)\,\big(\Phi_n^{\mathrm{s.c.}}( q')\big)^*\, G(\eta q',q;T)\;,
\end{equation}
when we insert the semiclassical expressions~\eqref{eq:vanvleck}, 
 we obtain a sum of integrals of the form
\begin{multline}
 \int\dmat q\,\dmat q' \frac{\BIGEXP{\displaystyle\frac{\imat}{\hbar} \int^q_{b_\beta} p_\beta(x,E_n)\,\dmat x}}{\sqrt{\partial_p H\big(p_\beta(q,E_n),q\big)}}
\frac{\BIGEXP{\displaystyle -\frac{\imat}{\hbar} \int^{q'}_{b_{\beta'}} p_{\beta'}(x,E_n)\,\dmat x}}{\sqrt{\partial_p H\big(p_{\beta'}(\eta q',E_n),q'\big)}}\times\\
\sqrt{\frac{\partial^2 S_\goo}{\partial q_i\partial q_f}
                \bigg|}_{(\eta q', q; T)}
 \EXP{\imat S_\goo(\eta q', q; T)/\hbar}\;.
\end{multline}
The stationary conditions
\begin{subequations}\label{eq:pprimeqqprime}
\begin{eqnarray}
   p_\beta(q,E_n)&=-&\partial_{q_i} S_\goo(\eta q', q; T)\;;\\
   \eta p_{\beta'}(q',E_n)&=& \partial_{q_f} S_\goo(\eta q', q; T)
\end{eqnarray}
\end{subequations}
select the classical trajectories~$\goo$ with energy~$E_n$ that go
from~$(p_i,q_i)=(p_\beta(q,E_n),q)$ at~$t(s_i)=0$ to ~$(p_f,q_f)=(\eta
p_{\beta'}(q',E_n),\eta q')$ at time~$t(s_f)=T$.  Then the value of
the exponent
\begin{equation}
  S_{\goo,\beta,\beta'}(E_n, T)=-E_n T+\tilde S_\goo(b_{\beta'}, b_{\beta^{}},E_n)
\end{equation}
depends only on the branches where the starting and ending points lie
and not on the precise location of these points on the branches.
Since~$q$ and~$q'$ correspond to the same torus, such a trajectory must
connect the two symmetric tori for~$\eta=-1$.  At a given~$E_n$
and~$T$, for a fixed~$q=q_i$ on the branch~$\beta$,
$q_f$ and~$\beta'$ are uniquely given and we can make the stationary
phase approximation for the integral on~$q'$.  Then if we insert
\eqref{eq:TEqiqf} at energy~$E_n$ into \eqref{eq:pprimeqqprime} and
differentiate it with respect to~$q$ or~$q'$, we obtain some
identities that, with \eqref{eq:SdEdT}, allow us to simplify the
combination of the prefactors and the remaining integral in~$q$ turns
out to be precisely of the form of the right hand side
of~\eqref{def:primitiveperiod}. \emph{A priori}, the domain of integration is included
in the domain of the branch~$\beta$ but is not necessarily equal to it
because when sliding  the starting point on the whole branch~$\beta$, the endpoint 
may cross a turning point and correspond to a jump of~$\beta'$.  However, we obtain characteristic times
that depend only on the geometry of the orbit, not on the number of
times the considered branch may be repeated as $s$ goes from~$s_i$ to
$s_f$.  As discussed in the case of the double well, 
if there exist different topological classes of~$\goo$, each of them being characterised
by an ordered sequence of turning points~$[\rho_1,\rho_2,\dots]$, we must add such contributions. Then, using
directly~$\tr\!\big(\hat\varPi_n\hat{U}(T)\big)\simeq 2\EXP{-\imat E_n
  T/\hbar}$, we have proven that
\begin{equation}\label{eq:scDeltangeneral}
  \Delta_n(T)\scl\frac{\hbar}{T}\!\!\!\sum_{[\rho_1,\rho_2,\dots]}\!\!\! (-1)^{\mu_\goo}A_{\beta,\beta'}(E_n,T)\,\EXP{\imat\tilde S_\goo(b_{\beta'},b_{\beta^{}},E_n)/\hbar}
\end{equation}
where the sum runs over all the sequences of turning points on the section~$\gos$ at energy~$E_n$
where one canonical variable is maintained real. There must exist for such a sequence, one half
symmetric orbit~$\goo$ starting on the branch~$\beta$ of the torus at energy~$E_n$,
crossing successively all the sequences~$[\rho_1,\rho_2,\dots]$ and ending on 
the branch~$\beta'$ at time~$T$.
 The dimensionless coefficients 
$A_{\beta,\beta'}(E_n,T)$ have a  $\hbar$-independent modulus of order one and depend only on the geometrical
properties of the branches.
  If some parts of the trajectory are repeated, their repetition numbers
do not appear in~$A_{\beta,\beta'}(E_n,T)$ but only in the cumulative quantities:
 the index~$\mu_\goo$ and 
the action $\tilde S_\goo(b_{\beta'}, b_{\beta^{}},E_n)$ given by
\begin{multline}\label{eq:Sbbetabbetaprime}
  \tilde S_\goo(b_{\beta'}, b_{\beta^{}},E_n)=\int^{s_f}_{s_i}p(s)\frac{\dmat q}{\dmat s}\,\dmat s\\
+\int_{b_{\beta^{}}}^{q(s_i)}p_{\beta^{}}(x,E_n)\,\dmat x
-\int_{b_{\beta'}}^{q(s_f)}p_{\beta'}(x,E_n)\,\dmat x\;.
\end{multline}
In the case of two branches, the computation of the coefficient can be 
 done exactly using the appropriate choice of phase conventions for the base 
points $b_\beta$ and~$A_{n,\beta}$; we obtain: 
\begin{equation}\label{eq:Abetabetap2branches}
  A_{\beta,\beta'}(E_n,T)=-1/2\;.
\end{equation}
We illustrate in the main body of this article, how to compute the sum in the right hand side 
of~\eqref{eq:scDeltangeneral}.


\end{document}